\documentclass[10pt]{article}
\usepackage{amsmath,amssymb,amsthm,amsopn,mathrsfs,mathdots,mathtools}
\usepackage{graphicx,verbatim,array,multicol,courier,dsfont,soul}
\usepackage[pdftex,bookmarks,colorlinks]{hyperref}
\usepackage{fullpage,anysize,relsize,color}
\usepackage[margin=1in]{geometry}
\usepackage[comma]{natbib}
\usepackage[section]{placeins}
\usepackage[labelformat=simple]{subfig}
\usepackage[applemac]{inputenc}
\usepackage{booktabs}
\usepackage[makeroom]{cancel}
\usepackage{enumerate}
\usepackage{enumitem}
\usepackage{float}

\newlist{inparaenum}{enumerate}{2}
\setlist[inparaenum,1]{label=(\roman*)}
\setlist[inparaenum,2]{label=(\roman{inparaenumi}\emph{\alph*})}

\newcommand{\der}{{\mathrm d}}

\newcommand{\bX}{{\boldsymbol{X}}}
\newcommand{\bx}{{\boldsymbol{x}}}
\newcommand{\bz}{{\boldsymbol{z}}}
\newcommand{\bb}{{\boldsymbol{b}}}
\newcommand{\bu}{{\boldsymbol{u}}}
\newcommand{\bw}{{\boldsymbol{w}}}
\newcommand{\diff}{\mathrm{d}}

\def\cL{\mathscr{L}}

\def\der{\text{d}}

\def\bmu{{\boldsymbol\mu}}
\def\bOmega{{\boldsymbol\Omega}}
\def\balpha{{\boldsymbol\alpha}}
\def\sbOmega{{\boldsymbol{\mathsmaller{\Omega}}}}
\def\bomega{{\boldsymbol\omega}}
\def\diag{\mbox{diag}}
\def\bzero{{\boldsymbol 0}}
\def\bM{{\boldsymbol M}}
\def\bba{{\boldsymbol a}}
\def\simp{\mathcal{S}_{\ds}}
\def\ds{{d}}
\def\bt{{\boldsymbol t}}
\def\bLambda{{\boldsymbol\Lambda}}
\def\blambda{{\boldsymbol\lambda}}
\def\prob{{\mathbb P}}
\def\nat{{\mathbb N}}

\newtheorem{theo}{Theorem}[section]

\newtheorem{cond}{Condition}
\newtheorem{prop}{Proposition}[section]

\newtheorem{lem}{Lemma}

\hypersetup{colorlinks,%
citecolor=blue,%
filecolor=green,%
linkcolor=red,%
urlcolor=magenta,%
}

\begin{document}

\title{Extremal properties of the multivariate extended skew-normal distribution}

\author{B. Beranger\footnote{School of Mathematics and Statistics, University of New South Wales, Sydney, Australia.}\;\,\footnote{Communicating Author: {\tt B.Beranger@unsw.edu.au}},\; S. A. Padoan\footnote{Department of Decision Sciences, Bocconi University of Milan, Italy.
},\; Y. Xu$^*$\; and\; S. A. Sisson$^*$
}

\date{}

\maketitle
\begin{abstract}
The skew-normal and related families are flexible and asymmetric parametric models suitable for modelling a diverse range of systems. 
We show that the multivariate maximum of a high-dimensional extended skew-normal random sample 
has asymptotically independent components and derive the speed of convergence of the  joint tail.
To describe the possible dependence among the components of the multivariate maximum, we show that under appropriate conditions an approximate multivariate extreme-value distribution that leads to a rich dependence structure can be derived.\\
\noindent Keywords: Asymptotic independence; Coefficient of upper-tail dependence; Pickands dependence function; Multivariate extreme-value distribution; Stable-tail dependence function.
\end{abstract}
%

%
%
%
\section{Introduction}\label{sec:intro}
The skew-normal and related families, 
such as the more flexible extended skew-normal and extended skew-$t$ distributions \citep[Ch 5.]{arellano2010, azzalini2014}, are suitable for data that exhibit
an asymmetric distribution, while still providing relatively simple probabilistic models.
For risk analysis in the fields of  insurance (credit risk management, loss 
ratios), climatology (floods, heat waves, storms) and health (influenza mortality),
it is of particular interest to study the tail behavior of the skew-normal and its related families
\citep[e.g.][Ch. 4]{peng2016, fung2014, liao2014, azzalini2014}.
As a consequence, a number of results on the limiting extreme-value distribution for the extremes of skew-normal and skew-$t$ samples have been obtained 
\citep[e.g.][]{chang2007, lysenko2009, padoan2011, beranger2017}.
However, while the extremal properties of skew-normal and skew-$t$ distributions have been extensively studied,
those of the more flexible extended skew-normal distribution have not yet been investigated.

In this contribution we derive the extremal properties of the multivariate extended skew-normal distribution. 
Recall that a $d$-dimensional random vector $\bX$ follows an extended skew-normal distribution \citep{arellano2010}, denoted 
as $\bX\sim ESN_d(\bmu, \bOmega, \balpha, \tau)$, if its probability density function (pdf) is given by
\begin{equation}
\label{eq:pdfESN}
\phi_d(\bx; \bmu, \bOmega, \balpha, \tau) = 
\frac{\phi_d(\bx;\bmu,\bOmega)}{\Phi\left(\tau/\sqrt{1+ Q_{\bar{\sbOmega}}(\balpha)} \right)} 
\Phi(\balpha^\top \bz + \tau), 
\quad \bx\in \real^d,
\end{equation}
where $\phi_d(\bx;\bmu,\bOmega)$ is a $d$-dimensional normal pdf with 
mean $\bmu \in \real^d$ and $d \times d$ 
covariance matrix $\bOmega$, $\bz = \bomega^{-1}(\bx-\bmu)$, 
$\bomega = \diag(\bOmega)^{1/2}$, 
$\bar{\bOmega} = \bomega^{-1} \bOmega \bomega^{-1}$, 
$Q_{\bar{\sbOmega}}(\balpha) = \balpha^\top \bar{\bOmega} \balpha$, 
$\Phi(\cdot)$ is the standard univariate normal cumulative distribution function (cdf) and
$\balpha \in \real^d$ and $\tau \in \real$ are the slant and  extension parameters, respectively, which control the nature of density deviations away from normality.
When $\tau = 0$ or $\tau=0$ and  $\balpha= \bzero$ the extended skew-normal distribution reduces to the skew-normal $SN_d(\bmu,\bOmega,\balpha)$ or the normal $N_d(\bmu,\bOmega)$ distribution. 
Without loss of generality, we work with location and scale standardised distributions throughout, so that  
$ESN_d(\bar{\bOmega},\balpha,\tau)$ and $\Phi_d(\bx; \bar{\bOmega}, \balpha, \tau)$
refer to the $d$-dimensional extended skew-normal distribution and extended skew-normal cdf
with location $\bmu=\bzero$ and correlation matrix $\bar{\bOmega}$, respectively. 
Finally, in the univariate setting, for brevity, we write the distributional parameters in the subscript of the pdf and cdf so that $\phi(x; \alpha,\tau)=\phi_{\alpha,\tau}(x)$ and $\Phi(x;\alpha,\tau)=\Phi_{\alpha,\tau}(x)$.

In this paper we establish that the multivariate maximum of a high-dimensional extended skew-normal random sample 
has asymptotically independent components. In particular, in the bivariate case we derive the speed of convergence 
of the joint upper tail. 
To describe the possible dependence between the components of the multivariate maximum,
we consider a similar approach to that introduced in \citet{husler1989}. We compute a multivariate maximum over a triangular array of extended skew-normal random vectors and, under suitable conditions, derive an approximate multivariate extreme-value distribution, for large sample sizes. This leads to a model with a rich extremal dependence structure, of which we  illustrate several features.

The paper is organized as follows. In Section~\ref{sec:back} we briefly review basic notions of 
multivariate extreme-value theory.
In Section~\ref{sec:esn_family} we show that the multivariate sample maximum has asymptotically independent components
and for the bivariate case deduce the convergence speed of the joint tail. We complete the Section by deriving an approximate multivariate extreme-value distribution and discuss some features of its extremal dependence structure.
All proofs are provided in the Appendix.

%
%
%
\section{Extreme-value theory background}\label{sec:back}

Let $I=\{1,\ldots,d\}$ be an index set denoting variables of interest.
Let $\bX_1, \ldots, \bX_n$, be a series of {\em iid} $d$-dimensional random vectors, where
$\bX_i = (X_{i,1}, \ldots, X_{i,d})^{\top}$ for $i=1,\ldots,n$, with
a continuous joint distribution function $F$ defined on $\real^d$, with marginal distributions $F_j$, $j\in I$.
The vector of ($n$-partial) sample maxima is defined componentwise as
$\bM_n = (M_{n,1}, \ldots, M_{n,d})^{\top}$ with $M_{n,j} = \max_{i=1, \ldots, n} X_{i,j},\quad j\in I.$
As with the univariate setting, if there is a sequence of normalising constants $\bba_n =(a_{n,1}, \ldots, a_{n,d})^\top > \bzero=(0,\ldots,0)^\top$ 
and  $\bb_n = (b_{n,1}, \ldots, b_{n,d})^\top \in \real^d$ such that 
\begin{align}
\label{eq:MultLimDist}
\lim_{n\to\infty}\Pr \left( \frac{\bM_n -\bb_n}{\bba_n} \leq \bx \right) 
= \lim_{n\to\infty} F^n(\bba_n \bx + \bb_n)=G(\bx),
\end{align}
for all continuity points $\bx=(x_1,\ldots,x_d)^\top\in\real^d$ of $G(\bx)$, and where $\bba_n\bx$ denotes componentwise multiplication, then if $G$ is a distribution function with nondegenerate margins
it is called a multivariate extreme-value distribution \cite[e.g.][Ch. 6]{beirlant2004}. 
Specifically, $G$ takes the form
$
G(\bx)=C\{G_1(x_1),\ldots,G_d(x_d)\},
$
$\bx\in\real^d$, where its univariate margins $G_j$, $j\in I$, are members of the GEV family \citep[e.g.][p. 47]{beirlant2004} and $C$ is an extreme-value
copula with expression
$$
C(\bu)=\exp\{-L(-\ln u_1,\cdots,-\ln u_d)\},\quad \bu\in (0,1]^d,
$$
where $\bu=(u_1,\ldots,u_d)^\top$ and where $L:[0,\infty)^d\mapsto[0,\infty)$ is the stable dependence function \citep[e.g.][Section~8.2.2]{beirlant2004}.
Specifically, 
\begin{equation}\label{eq:stable}
L( \bz ) = d\int_{\simp}\max\left(z_1w_1,\dots,z_dw_d\right)H(\diff \bw),\quad \bz\in[0,\infty)^d,
\end{equation}
where $\bw=(w_1,\ldots,w_d)^\top$, $\bz=(z_1,\ldots,z_d)^\top$ and where the angular measure $H$ is a probability measure defined on the $d$-dimensional unit simplex
$
\simp := \left\{ (v_1,\ldots, v_d) \in [0,1]^d: v_1+\cdots+v_d = 1 \right\}
$
satisfying the mean constraint $\int_{\simp} w_jH(\diff \bw)=1/d$ for all $j\in I$. 
By the homogeneity property of $L$ it follows that
\begin{equation*}
\label{eq:ellA}
  L( \bz ) = (z_1 + \cdots + z_d) \, A( \bt ), \qquad \bz \in [0, \infty)^d,
\end{equation*}
where $\bt=(t_1,\ldots,t_d)^\top$ with $t_j = z_j / (z_1 + \cdots + z_d)$ for $j = 1, \ldots, d-1$, 
$t_d = 1 - t_1 - \cdots - t_{d-1}$, 
where $A$ is Pickands dependence function \citep[e.g.][Section~8.2.5]{beirlant2004}, which is
the restriction of $L$ on $\simp$.
It quantifies the level of dependence between the  extremes, and
satisfies the condition $1/d\leq \max(t_1,\ldots,t_{d})\leq A(\bt)\leq 1$ for all $\bt\in\simp$, 
with the lower and upper bounds representing complete dependence and independence, respectively.

An important and useful summary of extremal dependence is the coefficient of upper-tail dependence, denoted by $\chi$ \citep[Ch. 2]{Li09,Joe97}.
In the bivariate case, it is constructed as the probability that $X_i$ and $X_j$, $i\neq j\in I$, are jointly extreme. Explicitly, 
$\chi :=\lim_{u\to0^+}\chi(u)$, where
\begin{equation}\label{eq:up_tail_fun}
\chi(u)=\frac{\Pr(F_i(X_i)\geq 1-u, F_j(X_j)\geq 1-u)}{u}, \quad u\in(0,1],
\end{equation}

\noindent where $0\leq\chi\leq1$. The variables $(X_i,X_j)$ are said to be asymptotically independent in the upper-tail
when $\chi=0$ and  are  asymptotically dependent when $\chi>0$. The case where $\chi=1$ represents
complete dependence between $X_i$ and $X_j$. On the basis of the speed of convergence of $\chi(u)$ to zero
as $u\to 0^+$, \citet{ledford1996} proposed
an approach to describe the sub-asymptotic, upper-tail dependence in the case of asymptotic independence. Specifically, they assumed that the
upper-tail dependence function $\chi(u)$ \eqref{eq:up_tail_fun} behaves as 
$
\chi(u)=u^{1/\eta-1}\, \cL(1/u), 
$
as $u \to 0^+$, where $\eta\in(0,1]$ is the coefficient of tail dependence and $\cL(1/u)$ is a  slowly varying function, such that $\cL(a/u)/\cL(u)\rightarrow 1$ as $u\to 0^+$, for fixed $a>0$.
Considering $\cL$ as a constant, at extreme levels margins 
are negatively associated when $\eta<1/2$, independent when $\eta=1/2$ and positively associated when $1/2<\eta< 1$. When
$\eta=1$ and $\cL(1/u)\nrightarrow0$  asymptotic dependence  is obtained.

%
%
%
\section{Extremes of extended skew-normal random samples}\label{sec:esn_family}
%
%
%

It is well known that the components of both normal and skew-normal random vectors 
are asymptotically independent. That is, the limit distribution of the normalised vector of componentwise maxima 
given by \eqref{eq:MultLimDist} is equal to the product of its marginal distributions \citep[e.g][pp. 285--87]{lysenko2009, beirlant2004}.
However, \citet{beranger2017} showed that for the skew-normal case, the rate of convergence to zero of the upper-tail dependence function $\chi(u)$ in \eqref{eq:up_tail_fun} depends on the slant parameters $\balpha$, and depending on the sign of the elements
of $\balpha$, this can occur at a faster or slower rate than that of the normal case.
Accordingly, from both theoretical and applied perspectives, it is important to understand whether 
 these results also hold for the tail behaviour of the extended skew-normal distribution, in which the extension parameter $\tau$ also plays a part in the speed of convergence.  We first consider the question of asymptotic dependence or asymptotic independence.
\begin{prop}[Asymptotic Independence]
\label{prop:asy_indep}
Let $\bX\sim ESN_d(\bar{\bOmega},\balpha,\tau)$.
Let $\chi(u)$ with $u\in(0,1]$ be the joint probability in \eqref{eq:up_tail_fun}.
Then, for every bivariate pair $(X_i,X_j)$ with $1\leq i<j\leq d$ we have that $\chi=0$.
\end{prop}
That is, it follows from Proposition \ref{prop:asy_indep}
that regardless of the degree of sub-asymptotic dependence, the components of the multivariate extended skew-normal distribution are asymptotically independent, and so the asymptotic distribution is a product of univariate standard Gumbel distributions.
We now examine the rate of convergence of $\chi(u)\rightarrow0$ in the extended skew-normal case. Here the primary aim is to evaluate the effect on the rate of convergence of the extension parameter $\tau$.
\begin{prop}[Bivariate Tail Convergence]
\label{prop:tail_conv}
Let $(X_1,X_2)\sim ESN_2(\bar{\bOmega},\balpha,\tau)$, where the off-diagonal term of $\bar{\bOmega}$ is $\omega\in[0,1)$, 
$\balpha\in\real^2$ and $\tau\in\real$. Set $K=\Phi(\tau/\sqrt{1+\alpha_1^2+\alpha_2^2+2\omega\alpha_1\alpha_2})$, 
$\bar{\alpha}_j=(1+\alpha_j^{*^{2}})^{1/2}$, $\alpha_j^{*}=(\alpha_j+\omega\alpha_{3-j})/
\{1+\alpha_{3-j}(1-\omega^2)\}^{1/2}$ for $j=1,2$.
Then, $\chi(u)\approx u^{1/\eta-1}\cL(1/u)$ as $u\to 0^+$, where
\begin{inparaenum}
\item \label{en:fcond_palpha} 
when either $\alpha_1,\alpha_2\geq0$ or $\omega>0$ and $\alpha_j\leq0$ and $\omega\alpha_{3-j}+\alpha_j\geq0$ for $j=1,2$, then
\begin{align*}
\eta=(1+\omega)/2,\quad
\cL(1/u)=(1+\omega)K/(1-\omega)(4\pi\ln(1/u))^{-\frac{\omega}{1+\omega}}.
\end{align*}
\item \label{en:scond_palpha} 
when $\omega>0$, 
$\alpha_j\leq0$ and $-\alpha_j\omega\leq \alpha_{3-j} < -\alpha_j/\omega$ for $j=1,2$, then
\begin{inparaenum}
\item \label{en:scond_palpha_sub_1} 
if $\alpha_{3-j}>-\alpha_j/\bar{\alpha}_j$, then
$$
\eta=\frac{(1-\omega^2)\bar{\alpha_j}^2}{1-\omega^2+(\bar{\alpha}_j^2-\omega)^2},\;
\cL(1/u)=\frac{\bar{\alpha}_j^2(1-\omega^2)K^{1/\eta-1/\{\bar{\alpha}_j(1+\omega)\}}}{(\bar{\alpha}_j-\omega)(1-\omega\bar{\alpha}_j)}(4\pi\ln(1/u))^{1/2\eta-1}.
$$
\item \label{en:scond_palpha_sub_2} 
if $\alpha_{3-j}<-\alpha_j/\bar{\alpha}_j$, then
\begin{align*}
\eta&=\left[\{1-\omega^2+(\bar{\alpha}_j^2-\omega)^2\}/\{(1-\omega)^2\bar{\alpha_j}^2\}+\left(\alpha_{3-j}-\alpha_j/\bar{\alpha}_j\right)^2\right]^{-1},\\
\cL(1/u)&=\frac{e^{-\tau^2/2}\bar{\alpha}_j^2(1-\omega^2)(\alpha_{3-j}-\alpha_j/\bar{\alpha}_j)^{-1} K^{1/\eta-1}}
{(\bar{\alpha}_j-\omega)\{1-\omega\bar{\alpha}_j+\alpha_{3-j}\alpha_j\bar{\alpha}_j(1-\omega^2)\}}(4\pi\ln(1/u))^{1/2\eta-3/2}.
\end{align*}
\end{inparaenum}
\item \label{en:tcond_nalpha} 
when either $\alpha_1,\alpha_2<0$ or $\omega>0$, $\alpha_j<0$ and  $0<\alpha_{3-j}<-\omega\alpha_j$ for $j=1,2$, then
\begin{align*}
\eta&=(1-\omega^2) \left\{\frac{\alpha_{3-j}^2(1-\omega^2)+1}{\bar{\alpha}_{3-j}^2}+\frac{\alpha_j^2(1-\omega^2)+1}{\bar{\alpha}_j^2}+\frac{2(\alpha_1\alpha_2(1-\omega^2)-\omega)}{\bar{\alpha}_1\bar{\alpha}_2}\right\}^{-1},\\
\cL(1/u)&=\frac{\bar{\alpha}_j^3\bar{\alpha}_{3-j}(1-\omega^2) \left[\{1-\omega\bar{\alpha}_{3-j}/\bar{\alpha}_j\}/(1-\omega^2)+
\alpha_j\left(\alpha_j+\alpha_{3-j}\bar{\alpha}_{3-j}/\bar{\alpha}_j\right)\right]^{-1}e^{-\tau^2/2}K^{1/\eta-1}}{[\{1+\alpha_{3-j}^2(1-\omega)\}\{\bar{\alpha}_j-\omega\bar{\alpha}_{3-j}\}
+\alpha_1\alpha_2(1-\omega^2)^{3/2}(1-\omega)\bar{\alpha}_{3-j}](\alpha_{3-j}\bar{\alpha}_j+\alpha_j\bar{\alpha}_{3-j})}\\
&\times (4\pi\ln(1/u))^{1/2\eta-3/2}.
\end{align*}
\end{inparaenum}
\end{prop}
From Proposition \ref{prop:tail_conv} we see that the contribution of the extension parameter $\tau$ to the rate of tail convergence is contained in the $K^\psi$ term, where the power $\psi$ is independent of $\tau$ and changes depending on the value of $\balpha$.
For a bivariate skew-normal distribution,  Figure \ref{fig:cut_dep} illustrates the behaviour of 
$\chi(1-v)$ against $v\to 1^{-}$, where $\chi(u)$ is the upper-tail dependence function \eqref{eq:up_tail_fun}, for different values of the model parameters $\omega$, $\alpha_1$, $\alpha_2$ and $\tau$. In each panel, for fixed $\omega$ and $\tau$, the speed of convergence of $\chi(1-v)$ to $0$ as $v\to 1^{-}$ is fastest when both slant parameters ($\alpha_1$, $\alpha_2)$ are negative. It is slower in any other case, with the slowest convergence rate depending on both the sign and magnitude of the slant parameters.
However the effect of $\tau$ on the rate of convergence is more straightforward. While fixing all other parameters, for lower values of $\tau$ (left panel) the rate of convergence is faster than for higher values (right panel).
%
%
%
\begin{figure}[t!]
	\centering
	\includegraphics[width=4.5cm,height=5cm,page=1]{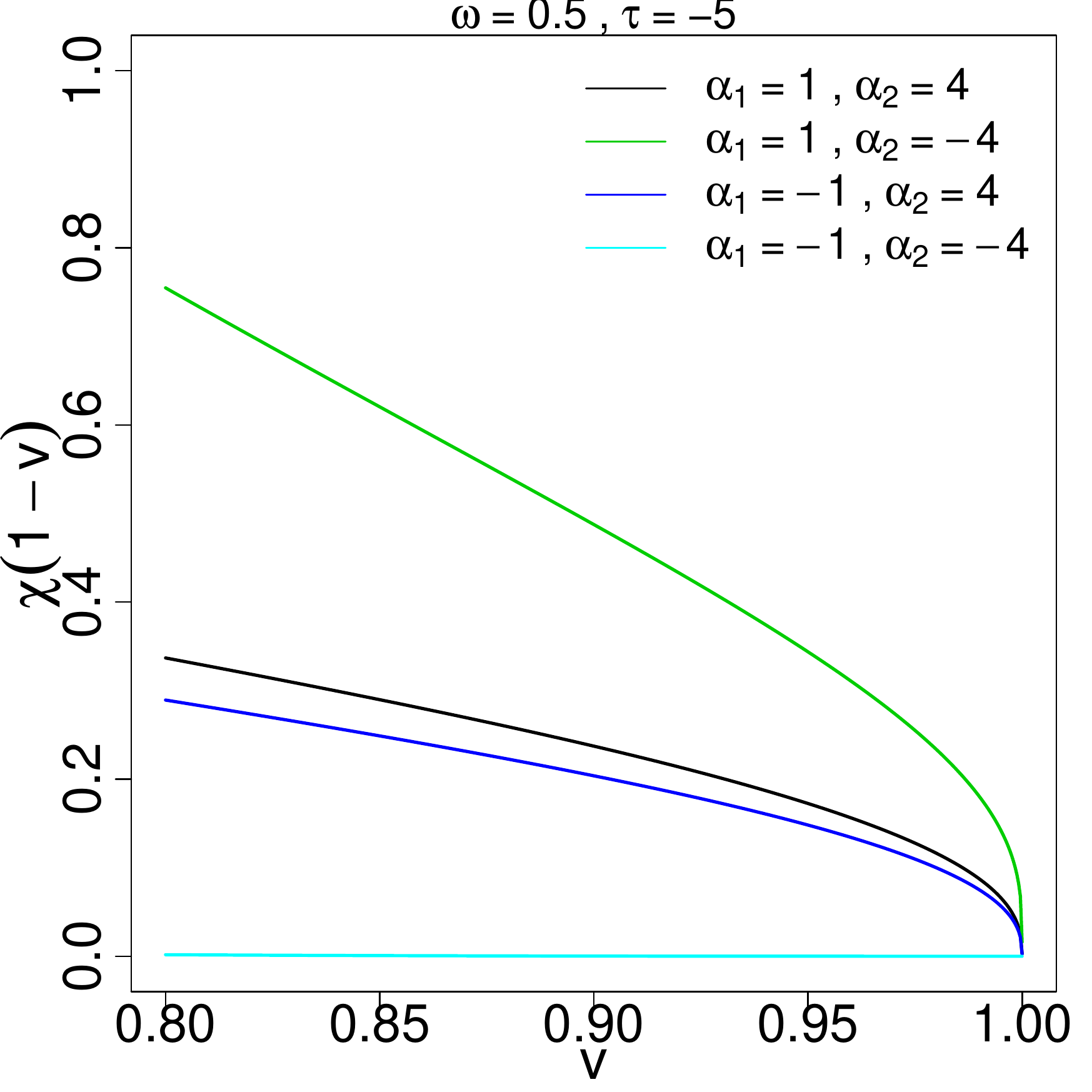}
	\includegraphics[width=4.5cm,height=5cm,page=2]{upper_tail_dep}
	\includegraphics[width=4.5cm,height=5cm,page=3]{upper_tail_dep}
	\caption{\small The behaviour of 
	$\chi(1-v)$ versus $v$ for different values of the parameters $\omega$, $\alpha_1$, $\alpha_2$ and $\tau$, for the bivariate skew-normal distribution. From left to right, the panels illustrate the effect of negative, zero and positive values of $\tau$, respectively.}
	\label{fig:cut_dep}
\end{figure}

Proposition \ref{prop:asy_indep} states that the marginal (componentwise) maxima 
$M_{n,1},\ldots, M_{n,d}$ are asymptotically independent, thereby determining an extremal framework that only permits independence among observed sample maxima. However, for data following Gaussian-type distributions, \citet{husler1989} developed an approach by which, under suitable conditions, an alternative non-independence asymptotic distribution for componentwise maxima may be formulated. This allows an extremal dependence structure possessing a rich class of asymptotic behaviour, ranging from independence to complete dependence,  to be derived. 
We now develop this alternative asymptotic distribution for the extend skew-normal class.

Precisely, for $n=1,2,\ldots$
let $\bX_{n,i}$, $i=1,\ldots,n$, be a triangular array of random vectors, where
$\bX_{n,i}=(X_{n,i;1},\ldots,X_{n,i;d})^\top$. Following  \citet{husler1989}, for each $n$, assume that $\bX_{n,1},\ldots,\bX_{n,n}$
are independent random vectors, where $\bX_{n,i}\sim ESN_d(\bar{\bOmega}_n,\balpha_n,\tau)$. 
Here, the dependence structure and asymmetry of the extended skew-normal distribution, as measured through $\bar{\bOmega}_n$ and $\balpha_n$, changes as the sample size $n$ increases. 
In particular it is assumed that the strength of dependence and asymmetry increase with $n$ at an appropriate rate. 
We formalise this as follows.
\begin{cond}\label{ass:tri_array}
For all $j\in I$, the elements of 
$\balpha_n=(\alpha_{n;1},\ldots,\alpha_{n;d})^\top$ satisfy $\alpha_{n;j}\to\pm\infty$ as $n\to\infty$ and
$$
\alpha_j^{\circ}=\lim_{n\to\infty} \alpha_{n;j}(\ln n)^{-1/2} \in\real,
$$
with $\alpha_1^{\circ}+\cdots+\alpha_d^{\circ}=0$. For every $i,j\in I$,  
the correlations $\omega_{n;i,j}$ of the $d$-dimensional matrix $\bar{\bOmega}_n$ satisfy
$$
\lambda^2_{i,j}= \lim_{n\to\infty}(1-\omega_{n;i,j}) \ln n \in(0,\infty].
$$
\end{cond}
Under the assumptions in Condition \ref{ass:tri_array}, we are now able to establish  \citet{husler1989}'s alternative extremal limit in the case of the extended skew-normal distribution.
\begin{theo}\label{theo:tri_array}
Consider a triangular array  of extended skew-normal random vectors $\bX_{1,n},\ldots,\bX_{n,n}$, $n=1,2,\ldots$.
Let $\bM_{n,n}=(M_{n,n;1},\ldots,M_{n,n;d})^{\top}$ where 
$
M_{n,n;j}=\max(X_{n,1;j},X_{n,2;j},\ldots,X_{n,n;j}), 
$
$j\in I$.
Under the assumptions in Condition \ref{ass:tri_array} there are sequences of normalising constants $\bba_n>\bzero$
and $\bb_n\in\real^d$ such that $\Phi_d^n(\bba_n\bx+\bb_n;\bar{\bOmega}_n,\balpha_n,\tau)\to G(\bx)$ 
as $n\to\infty$, where the univariate margins of $G$ are standard Gumbel distributions, i.e.~$G(x)=e^{-e^{-x}}$ with $x\in\real$, and
\begin{equation}\label{eq:skew_HR}
L(\bz)=\sum_{j=1}^d z_j \, \Phi_{d-1}
\left\{
\left(
\lambda_{ij} + \frac{1}{2 \lambda_{ij}} \log \frac{\tilde{z}_j}{\tilde{z}_i},i \in I_j\right)^{\top}; 
\bar{\bLambda}_j,\tilde{\balpha}_j,\tilde{\tau}_j
\right\},\, \bz \in [0,\infty)^d,
\end{equation}
and where $\bar{\bLambda}_j$ is a $(d-1)\times(d-1)$ correlation matrix with upper diagonal entries
$
\frac{\lambda^2_{ij} + \lambda^2_{kj} - \lambda^2_{ik}}{2\lambda_{ij}\lambda_{kj}},\quad j\in I,\;i,k\in I_j=I \backslash\{j\} 
$
$\tilde{\balpha}_j=(\sqrt{2}\,\alpha_{i}^{\circ}\lambda_{i,j},i\in I_j)^{\top}$, 
$\tilde{\tau}_j=\tau-\sum_{i\in I_j}\sqrt{2}\,\alpha_i^{\circ}\lambda_{i,j}$ and 
$$
\tilde{z}_i=z_i\Phi\left(\frac{\tau-\sum_{k\in I_j} \sqrt{2}\lambda^2_{k,j}\alpha^{\circ}_k}
{\sqrt{1+\sum_{k,m \in I_j} \alpha^{\circ}_k \alpha^{\circ}_m (\lambda^2_{k,j} + \lambda^2_{m,j}-\lambda^2_{k,m})}}\right)/
\Phi\left(\frac{\tau}
{\sqrt{1+\sum_{k,m \in I_j} \alpha^{\circ}_k \alpha^{\circ}_m (\lambda^2_{k,j} + \lambda^2_{m,j}-\lambda^2_{k,m})}}\right)
$$
and $\tilde{z}_j$ is defined as $\tilde{z}_i$ but where the index $i$ is replaced by $j$ and vice versa.
\end{theo}
For the resulting multivariate extreme-value distribution in Theorem \ref{theo:tri_array} we may derive  representations
of the extremal dependence.
In particular, from \eqref{eq:skew_HR} we may construct Pickands dependence function as
\begin{equation}\label{eq:pickands}
A(\bt)=\sum_{j=1}^d t_j \, \Phi_{d-1}
\left\{
\left(
\lambda_{ij} + \frac{1}{2 \lambda_{ij}} \log \frac{\tilde{t}_j}{\tilde{t}_i},i \in I_j\right)^{\top}; 
\bar{\bLambda}_j,\tilde{\balpha}_j,\tilde{\tau}_j
\right\},
\end{equation}
for $\bt=(t_1,\ldots,t_d)^\top$, 
where $\tilde{t}_j$ and $\tilde{t}_i$ are defined as $\tilde{z}_j$ and $\tilde{z}_i$. 

By exploiting the method described in \citep[e.g.][pp. 263-264, 292-293]{beirlant2004} the angular measure $H$ (defined through \eqref{eq:stable}) 
relative to \eqref{eq:skew_HR} may be derived. Specifically, $H$ places mass only in the interior of the simplex and so the angular density on $\simp$ may be expressed as
\begin{equation}\label{eq:angular_den}
h(\bw)=\frac{\phi_{d-1}
\left\{
\left(
\lambda_{i1} + \frac{1}{2 \lambda_{i1}} \log \frac{\tilde{w}_i}{\tilde{w}_1},i \in I_1\right)^{\top}; 
\bar{\bLambda}_1,\tilde{\balpha}_1,\tilde{\tau}_1
\right\}}{2\,w_1^2 \prod_{i=2}^d w_i\lambda_{i,1}},\quad \bw\in\simp,
\end{equation}
where $\tilde{w}_j$ and $\tilde{w}_i$ are defined as $\tilde{z}_j$ and $\tilde{z}_i$. 
Finally, for a bivariate random vector $(Z_1,Z_2)$ with  distribution given in Theorem \ref{theo:tri_array}, the coefficient upper-tail dependence in \eqref{eq:up_tail_fun} is
\begin{equation}\label{eq:cut_dep}
\begin{split}
\chi&=1-\Phi
\left(
\lambda_{1,2} + \frac{1}{2 \lambda_{1,2}} \log \frac{\Phi\left(\frac{\tau-\sqrt{2}\lambda_{1,2}^2\alpha^{\circ}_2}{1+2\lambda_{1,2}^2\alpha^{\circ}_1}\right)}{\Phi\left(\frac{\tau+\sqrt{2}\lambda_{1,2}^2\alpha_1^{\circ}}{1+2\lambda_{1,2}^2\alpha^{\circ}_2}\right)}; 
-\sqrt{2}\lambda_{1,2}\alpha_2^{\circ},\tau+\sqrt{2}\lambda_{1,2}^2\alpha_2^{\circ}
\right)\\
&= 1-\Phi
\left(
\lambda_{1,2} + \frac{1}{2 \lambda_{1,2}} \log \frac{\Phi\left(\frac{\tau+\sqrt{2}\lambda_{1,2}^2\alpha^{\circ}_1}{1+2\lambda_{1,2}^2\alpha^{\circ}_2}\right)}{\Phi\left(\frac{\tau-\sqrt{2}\lambda_{1,2}^2\alpha^{\circ}_2}{1+2\lambda_{1,2}^2\alpha^{\circ}_1}\right)}; 
\sqrt{2}\lambda_{1,2}\alpha^{\circ}_1,\tau-\sqrt{2}\lambda_{1,2}^2\alpha^{\circ}_1
\right).
\end{split}
\end{equation}
%
%
%
\begin{figure}[H]
	\centering
	\includegraphics[width=0.26\textwidth,page=1]{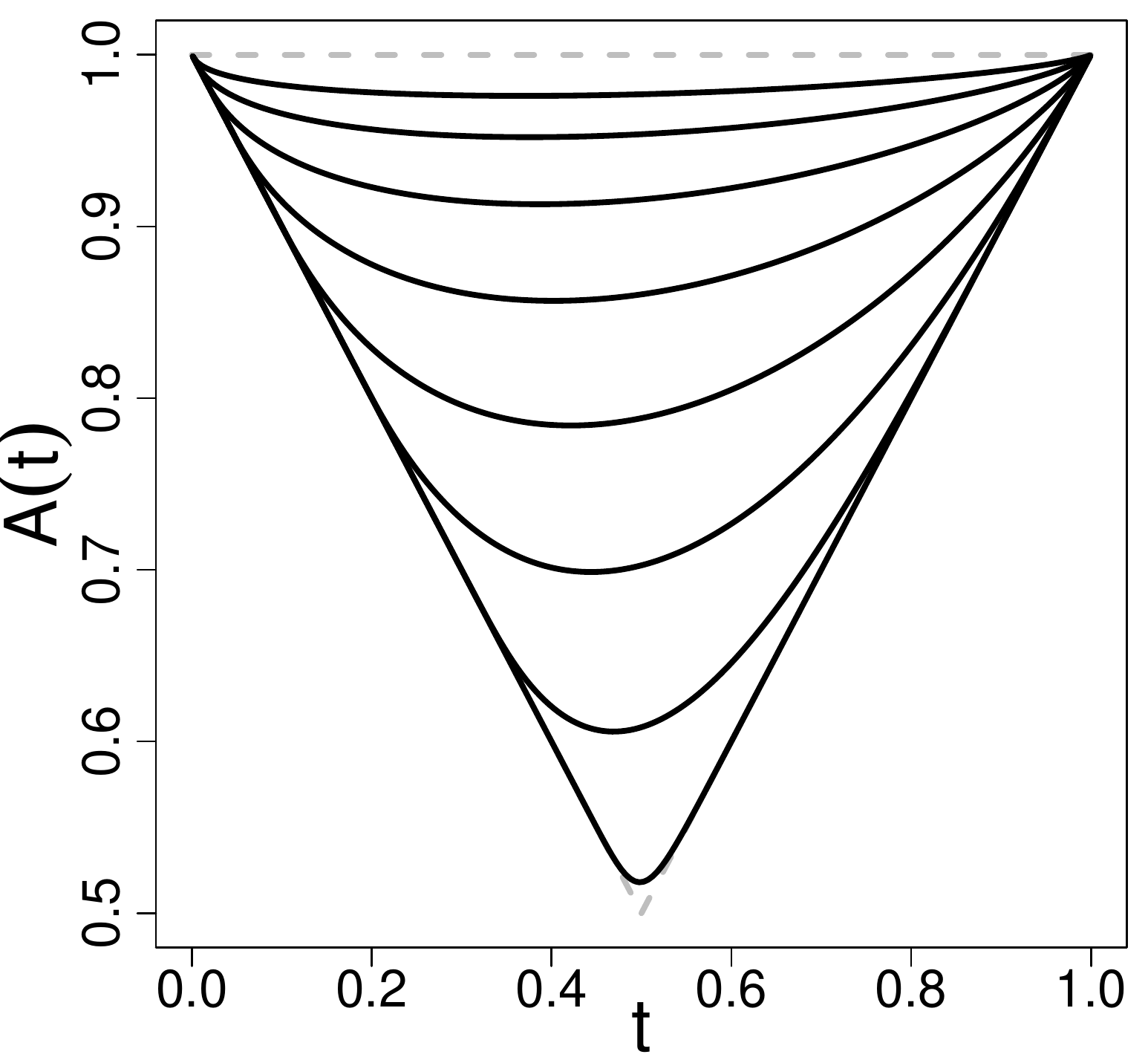}
	\includegraphics[width=0.26\textwidth,page=2]{pickands}
	\includegraphics[width=0.26\textwidth,page=3]{pickands}\\
	\includegraphics[width=0.26\textwidth,page=1]{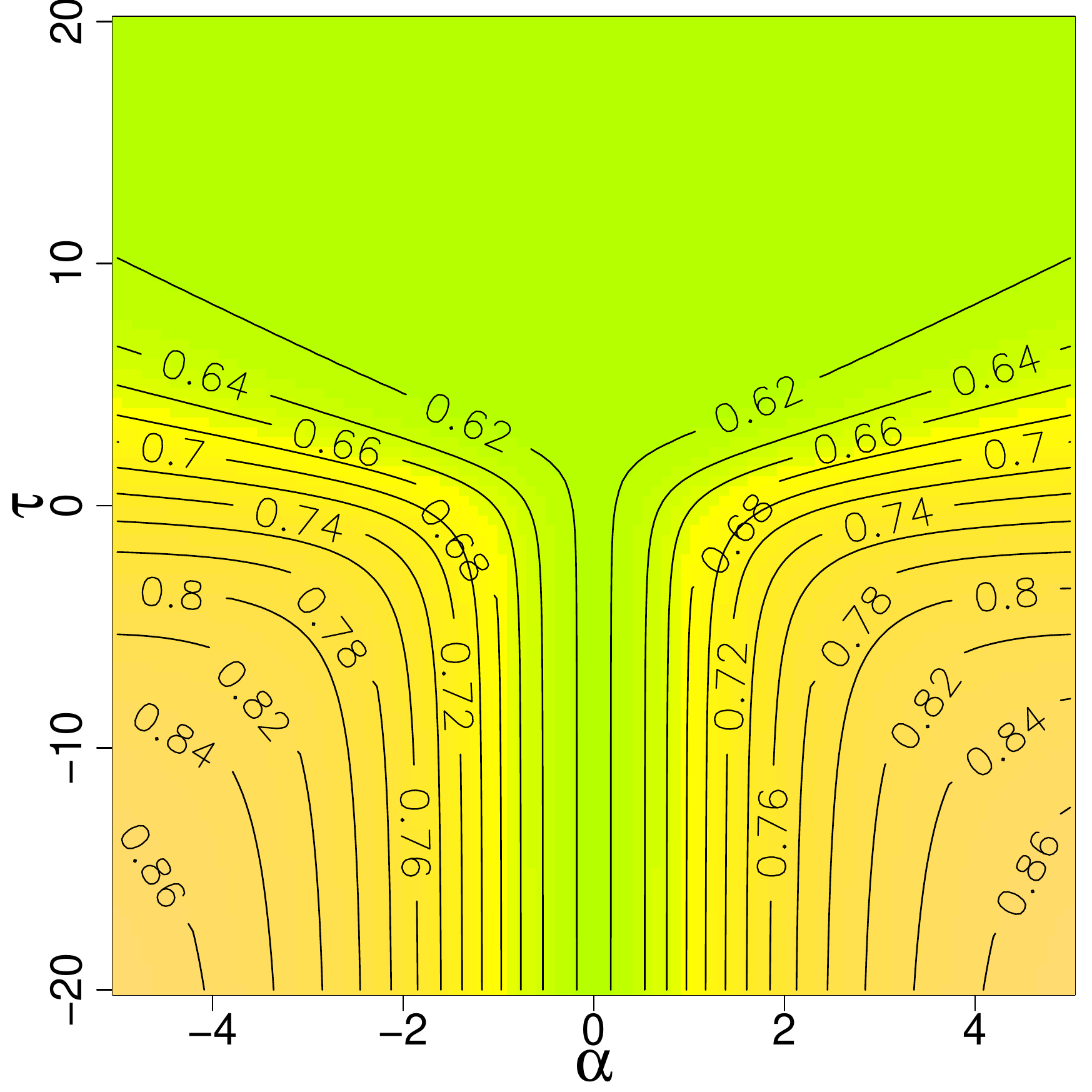}
	\includegraphics[width=0.26\textwidth,page=2]{Chiup_SkewHR}
	\includegraphics[width=0.26\textwidth,page=3]{Chiup_SkewHR}\\
	\includegraphics[width=0.26\textwidth,page=1]{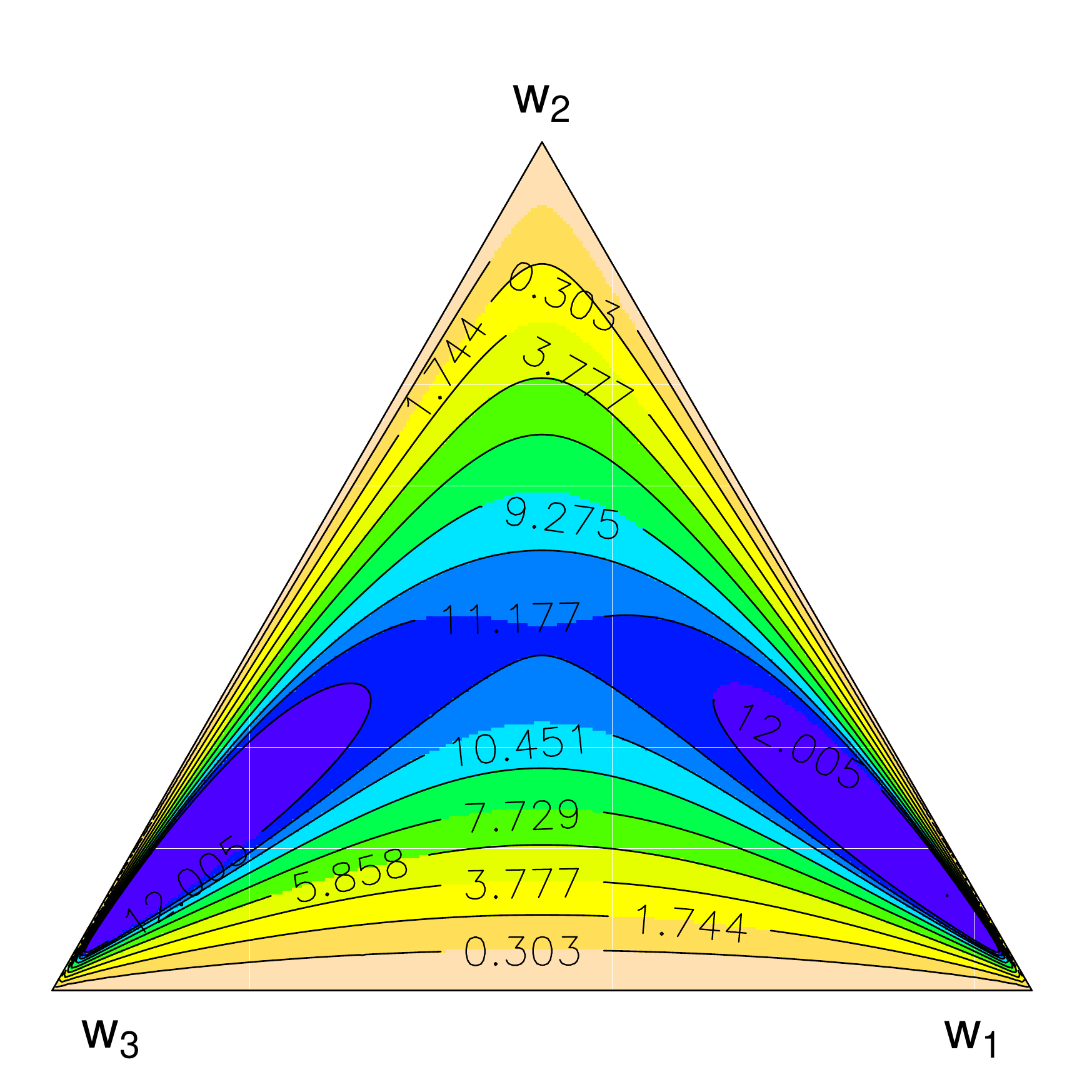}
	\includegraphics[width=0.26\textwidth,page=2]{SkewHR}
	\includegraphics[width=0.26\textwidth,page=3]{SkewHR}\\
	\includegraphics[width=0.26\textwidth,page=4]{SkewHR}
	\includegraphics[width=0.26\textwidth,page=5]{SkewHR}
	\includegraphics[width=0.26\textwidth,page=6]{SkewHR}
	\caption{\small Extremal dependence for the extended skew-normal distribution: (Top panels) Pickands dependence function, $A(\bt)$, (second row) the coefficient of upper-tail dependence $\chi$ and (bottom two rows) the
	angular density, $h(\bw)$, for different values of $\alpha^\circ, \tau$ and $\lambda$ (see main text for details).} 
	\label{fig:cut_dep}
\end{figure}
%
%
Figure \ref{fig:cut_dep} graphically illustrates a range of extremal dependence structures in terms of Pickands dependence function $A(\bt)$ \eqref{eq:pickands}, the angular density $h(\bw)$ \eqref{eq:angular_den} and the coefficient of upper tail dependence $\chi$ \eqref{eq:cut_dep}.
Each bivariate Pickands dependence function (top row) is constructed with $\lambda$ taking eight equally spaced values between $0.1$ and $3$.
Left to right, the top panels illustrate left-skewed (with $\alpha^{\circ}=-20$ and $\tau=-6$), symmetric ($\alpha^{\circ}=\tau=0$) and right-skewed ($\alpha^{\circ}=20$, $\tau=6$) dependence functions.
The bivariate coefficient of upper-tail dependence (second row), is illustrated for different
values of $\alpha^{\circ} \in [-5,5]$ and $\tau \in [-20,20]$ and, from left to right, with $\lambda=0.5,1$ and $2.5$. It is apparent that for fixed values of $\alpha^{\circ}$ and
$\tau$, $\chi$ decreases for increasing values of the dependence parameter 
$\lambda$. Similarly, for fixed values of $\lambda$ and $\alpha^{\circ}$, $\chi$ increases
for decreasing values of $\tau$. 
Finally, for fixed values of $\lambda$ and $\tau$,
 $\chi$ increases for increasing values of $|\alpha^{\circ}|$.

The bottom two rows illustrate trivariate angular densities for  
$\blambda^\top=(\lambda_{12}, \lambda_{13}, \lambda_{23})=(0.52, 0.71, 0.52)$ and different
values of the parameters 
$\balpha^\circ=(\alpha_1^{\circ},\alpha_2^{\circ},\alpha_3^{\circ})^\top$ and $\tau$. 
From left to right and top to bottom, the plots are produced with
slant parameters $\balpha$ given by $(0,0,0)$, $(0,5,-5)$, $(-5,5,0)$, $(4,-7,3)$, $(6,0,-6)$ and
$(6,-3,3)$ and extension parameters $0$, $0$, $0$, $3$, $5$ and $-5$, respectively.
The mass in the left panel of the third row concentrates around the centre of the simplex
meaning that there is strong dependence among all variables.
In the middle (and right) panels of the third row, the mass is concentrated on the bottom left (right) corner and left (right) edge. This means that two variables are themselves mildly dependent, and weakly
dependent of the third. 
In the bottom row, the mass in the left panel is concentrated on one corner
and two edges, meaning that one variable is mildly dependent on the other two, and these are weakly dependent
themselves. 
In the centre panel of the fourth row, 
the mass concentration in the centre panel is in the centre and on two edges, meaning
that one variable is strongly dependent on the others, and these are themselves weakly dependent. 
Finally, in the right panel the mass concentrates on one edge, so that two variables are
strongly dependent but they are each weakly dependent on the third.

\section{Summary}

The success of the multivariate skew-normal family is also due to its stochastic representations which motivate its use as stochastic model for data. For instance, sampling from the multivariate extended skew-normal distribution can be achieved through the distribution of the first $d$ components of a $(d+1)$-dimensional Gaussian random vector, conditionally that the $(d+1)$th component satisfies a certain condition \citep[][Ch.~5.1.3, 5.3.3]{arellano2010, azzalini2014}.
We have studied the extremal behaviour of extended skew-normal random vectors. Although their multivariate sample maximum has  asymptotically independent components, we have showed that the slant and extension parameters
affect the speed of convergence  of the joint upper tail for each of its bivariate components. Furthermore, we have derived the asymptotic
distribution for the sample maximum of a  
triangular array of independent extended skew-normal random vectors, under appropriate conditions on the correlations and the slant parameters. 
This produces a skewed version of the well-known H\"{u}sler-Reiss model \citep{husler1989}, where the skewness of such a distribution is affected by the extension parameter. \citet{hashorva+l16} have investigated the asymptotic
distribution for the sample maximum of a triangular array of independent bivariate skew-elliptical triangular arrays. 
They have also found that a modified version of the  H\"{u}sler-Reiss model emerges as possible asymptotic distribution under an appropriate condition on the random radius relative to the elliptical random vectors.
Their result differ from our result. In future would be interesting to investigate the extremal properties of a multivariate skew-elliptical distributions \citep[][Ch.~6]{azzalini2014} and study the relation with the triangular array type of approach investigated in \citet{hashorva+l16}.

\section*{Acknowledgments}

We thank two anonymous referees for having carefully read the first version of this manuscript and for their helpful comments that have contributed to improving the presentation of this article. 
BB and SAS are supported by the Australian Centre of Excellence in Mathematical and Statistical Frontiers (ACEMS; CE140100049) and by the Australian Research Council Discovery Projects Scheme (FT170100079). SAP is supported by the Bocconi Institute for Data Science and Analytics (BIDSA).

%
%
%
%
\appendix
\section{Proofs}\label{sec:appendix}

Auxiliary proofs  and details of the results  in Lemmas \ref{lem:common_step}
--\ref{lem:second_step} are  provided in the Supplementary Material.

\subsection{Proof of Proposition~\ref{prop:asy_indep}}
For $1\leq i<j\leq d$ we define
\begin{equation}\label{eq:marginal_par}
\alpha_i^{\ast}=(\alpha_i+\omega_{i,j}\alpha_j)/\{1+\alpha_j^2(1-\omega^2_{i,j})\}^{1/2},\quad
\alpha_j^{\ast}=(\alpha_j+\omega_{i,j}\alpha_i)/\{1+\alpha_i^2(1-\omega^2_{i,j})\}^{1/2}.
\end{equation}
We analyse the following four possible scenarios: (a) $0<\alpha_i^{\ast}<\alpha_j^{\ast}$,  (b)
$\alpha_i,\alpha_j<0$ and assume  $\alpha_i^{\ast}<\alpha_j^{\ast}$ with
$\alpha_i^{\ast}<0$, (c) $\alpha_i^{\ast},\alpha_j^{\ast}<0$ and assume that $\alpha_i<0$ and
$\alpha_j\geq 0$
and (d) $\alpha_i^{\ast}<0$ and $\alpha_j^{\ast}\geq 0$. Interchanging $\alpha_i^{\ast}$ with $\alpha_j^{\ast}$ produces the same results.
For brevity we set $\bar{\alpha}_i=(1+\alpha_i^{\ast 2})^{1/2}$ and 
$\bar{\alpha}_j=(1+\alpha_j^{\ast 2})^{1/2}$.
We first need the following result.
\begin{lem}\label{lem:common_step}
Let $\bX\sim ESN_d(\bar{\bOmega},\balpha,\tau)$.
for every pair $(X_i,X_j)$ with $1\leq i<j\leq d$ we have that
under the scenarios (a) and (b) 
$$
\lim_{x\to\infty}\frac{\Pr(X_j\geq x, X_i\geq x)}{\Pr(X_i\geq x)}=0.
$$
While under the scenario (c) and (d) we respectively have 
$$
\lim_{x\to\infty}\frac{\Pr\left(X_j\geq x, X_i\geq x\bar{\alpha}_j/\bar{\alpha}_i\right)}
{\Pr\left(X_i\geq x\bar{\alpha}_j/\bar{\alpha}_i\right)}=0 \quad \mbox{and}\quad
\lim_{x\to\infty}\frac{\Pr\left(X_j\geq x, X_i\geq x/\bar{\alpha}_i\right)}
{\Pr\left(X_i\geq x/\bar{\alpha}_i\right)}=0.
$$
\end{lem}

Consider the case (a) $0<\alpha_i^{\ast}<\alpha_j^{\ast}$. Using definition \eqref{eq:marginal_par} this assumption implies the inequality
$$
\{1+\alpha_i^2(1-\omega_{i,j}^2)\}(\alpha_j+\omega_{i,j}\alpha_j)^2<(\alpha_j+\omega_{i,j}\alpha_i)^2\{1+\alpha_j^2(1-\omega_{i,j}^2)\}
$$ 
and from this with elementary computations we obtain 
$\alpha_i^2<\alpha_j^2$ and 
$$
\tau_j^{\ast}=\tau/\{1+\alpha_i^2(1-\omega_{i,j}^2)\}^{1/2}>\tau/\{1+\alpha_j^2(1-\omega_{i,j}^2)\}^{1/2}=\tau_i^{\ast}.
$$
Therefore, as $x \to\infty$,
$$
\frac{\phi(x)\Phi(\alpha_i^{\ast}x+\tau_i^{\ast})}
{\Phi\left(\tau_i^{\ast}/\bar{\alpha}_i\right)}<
\frac{\phi(x)\Phi(\alpha_j^{\ast}x+\tau_j^{\ast})}
{\Phi\left(\tau_j^{\ast}/\bar{\alpha}_j\right)},
$$
which implies that $1-\Phi_{\alpha_i^{\ast},\tau_i^{\ast}}(x)< 1-\Phi_{\alpha_j^{\ast},\tau_j^{\ast}}(x)$
and $\Phi_{\alpha_i^{\ast},\tau_i^{\ast}}(x) > \Phi_{\alpha_j^{\ast},\tau_j^{\ast}}(x)$ as
$x\to\infty$. 
Then
\begin{equation*}\label{eq:asym_ind_ineq}
\begin{split}
\chi&=\lim_{u\to0^+}\Pr(\Phi_{\alpha_j^{\ast},\tau_j^{\ast}}(X_j)\geq 1-u | \Phi_{\alpha_i^{\ast},\tau_i^{\ast}}(X_i)\geq 1-u)\\
&=\lim_{x\to\infty}\Pr(\Phi_{\alpha_j^{\ast},\tau_j^{\ast}}(X_j)\geq \Phi_{\alpha_i^{\ast},\tau_i^{\ast}}(x) | X_i\geq x)\\
&\leq \lim_{x\to\infty}\Pr(\Phi_{\alpha_j^{\ast},\tau_j^{\ast}}(X_j)\geq \Phi_{\alpha_j^{\ast},\tau_j^{\ast}}(x) | X_i\geq x)
=\lim_{x\to\infty}\frac{\Pr(X_j\geq x, X_i\geq x)}{\Pr(X_i\geq x)}.
\end{split}
\end{equation*}
By Lemma \ref{lem:common_step} the last limit is equal to zero and therefore $\chi=0$.
 
Consider  case (b) $\alpha_i,\alpha_j<0$ and assume $\alpha_i^{\ast}<\alpha_j^{\ast}$ with
$\alpha_i^{\ast}<0$. Using similar arguments we obtain
$$
\chi < \lim_{x\to\infty}\frac{\Pr(X_j\geq x, X_i\geq x)}{\Pr(X_i\geq x)}.
$$
Then, by applying Lemma \ref{lem:common_step} we obtain $\chi=0$.

Consider  case (c) $\alpha_i^{\ast},\alpha_j^{\ast}<0$ and assume that $\alpha_i<0$ and
$\alpha_j\geq 0$, which implies that $\alpha_i^{\ast}<\alpha_j^{\ast}$. 
Applying Proposition 2.2 from \citet{boris+p+x+s2018} to $1-\Phi_{\alpha_i^{\ast},\tau_i^{\ast}}(x)$ and 
 Mill's ratio \citep{mills1926} to $\Phi(\alpha_i^{\ast}x+\tau_i^{\ast})$ we obtain
\begin{equation*}
\begin{split}
1-\Phi_{\alpha_i^{\ast},\tau_i^{\ast}}(x)&\approx \frac{\phi(x)\Phi(\alpha_i^{\ast}x+\tau_i^{\ast})}
{\Phi\left(\tau_i^{\ast}/\bar{\alpha}_i\right)\{\bar{\alpha}_i^2 x+\alpha_i^{\ast}\tau_i^{\ast}\}}\quad\mbox{as } x\to\infty\\
&\approx \frac{\phi(x)\phi(\alpha_i^{\ast}x+\tau_i^{\ast})}
{\Phi\left(\tau_i^{\ast}/\bar{\alpha}_i\right)\{\bar{\alpha}_i^2 x+\alpha_i^{\ast}\tau_i^{\ast}\}\{-(\alpha_i^{\ast}x+\tau_i^{\ast})\}}\quad\mbox{as } x\to \infty\\
&\approx\frac{\phi\left(x\bar{\alpha}_i\right)}{\Phi\left(\tau_i^{\ast}/\bar{\alpha}_i\right)\bar{\alpha}_i^2(-\alpha_i^{\ast})\sqrt{2\pi}x^2}\quad\mbox{as } x\to\infty.
\end{split}
\end{equation*}
Now note that
\begin{equation*}
\begin{split}
\phi\left(x(\bar{\alpha}_j/\bar{\alpha}_i)\bar{\alpha}_i\right)&=\phi\left(x\bar{\alpha}_j\right)\\
\frac{\phi\left(x(\bar{\alpha}_j/\bar{\alpha}_i)\bar{\alpha}_i\right)}{\Phi\left(\tau_j^{\ast}/\bar{\alpha}_j\right)(\bar{\alpha}_j^2/\bar{\alpha}_i^2)\bar{\alpha}_i^2(-\alpha_j^{\ast})\sqrt{2\pi}x^2}
&=\frac{\phi\left(x\bar{\alpha}_j\right)}{\Phi\left(\tau_j^{\ast}/\bar{\alpha}_j\right)\bar{\alpha}_j^2(-\alpha_j^{\ast})\sqrt{2\pi}x^2}.
\end{split}
\end{equation*}
Since $\alpha_i^{\ast}<\alpha_j^{\ast}$ then $-1/\alpha_i^{\ast}<-1/\alpha_j^{\ast}$ and it follows that
\begin{equation*}
\begin{split}
\frac{\phi\left(x(\bar{\alpha}_j/\bar{\alpha}_i)\bar{\alpha}_i\right)}{\Phi\left(\tau_j^{\ast}/\bar{\alpha}_j\right)(\bar{\alpha}_j^2/\bar{\alpha}_i^2)\bar{\alpha}_i^2(-\alpha_i^{\ast})\sqrt{2\pi}x^2}
&<\frac{\phi\left(x\bar{\alpha}_j\right)}{\Phi\left(\tau_j^{\ast}/\bar{\alpha}_j\right)\bar{\alpha}_j^2(-\alpha_j^{\ast})\sqrt{2\pi}x^2}.
\end{split}
\end{equation*}
Therefore, $1-\Phi_{\alpha_i^{\ast},\tau_i^{\ast}}(x\bar{\alpha}_j/\bar{\alpha}_i)<1-\Phi_{\alpha_j^{\ast},\tau_j^{\ast}}(x)$
and $\Phi_{\alpha_j^{\ast},\tau_j^{\ast}}(x)< \Phi_{\alpha_i^{\ast},\tau_i^{\ast}}(x\bar{\alpha}_j/\bar{\alpha}_i)$ and $\Phi_{\alpha_i^{\ast},\tau_i^{\ast}}(x\bar{\alpha}_j/\bar{\alpha}_i)<\Phi_{\alpha_i^{\ast},\tau_i^{\ast}}(x)$.
From this, with some manipulation we may obtain
\begin{equation*}\label{eq:chi_asym}
\chi \leq \lim_{x\to\infty}\frac{\Pr\left(X_j\geq x, X_i\geq x\bar{\alpha}_j/\bar{\alpha}_i\right)}
{\Pr\left(X_i\geq x\bar{\alpha}_j/\bar{\alpha}_i\right)}.
\end{equation*}
Now, applying Lemma \ref{lem:common_step} we obtain $\chi=0$.

Finally, consider  case (d) $\alpha_i^{\ast}<0$ and $\alpha_j^{\ast}\geq 0$.
Note that as $x\to\infty$ we have 
$$
\Phi\left(\alpha_j^{\ast}\bar{\alpha}_ix+\tau^{\ast}\right)>
\frac{1}{\sqrt{2\pi}(-\alpha_i^{\ast})\bar{\alpha}_jx}
$$
which implies that 
$1-\Phi_{x \alpha_j^{\ast},\tau_{j}^{\ast}}\left(\bar{\alpha}_i\right)>
1-\Phi_{\alpha_i^{\ast},\tau_{i}^{\ast}}(x)$ and $\Phi_{\alpha_j^{\ast},\tau_{j}^{\ast}}\left(x \bar{\alpha}_i\right)<
\Phi_{\alpha_i^{\ast},\tau_{i}^{\ast}}(x)$ as $x\to\infty$. These results imply 
that
\begin{align*}
\chi\leq  \lim_{x\to\infty}\frac{\Pr\left(X_j\geq x, X_i\geq x/\bar{\alpha}_i\right)}
{\Pr\left(X_i\geq x/\bar{\alpha}_i\right)}.
\end{align*}
Then, by applying Lemma \ref{lem:common_step} we obtain $\chi=0$.
Since $\chi=0$ for all $1\leq i<j\leq d$ then by \citet[Proposition 5.27]{resnick1987} 
we have that $\bX\sim ESN_d(\bar{\bOmega},\balpha,\tau)$ has asymptotically independent components.

%

\subsection{Proof of Proposition~\ref{prop:tail_conv}}

From \citet{arellano2010}, recall that if $\bX\sim ESN_2(\bar{\bOmega},\balpha,\tau)$ then for $j=1,2$ we have
$$
X_j\sim ESN(\alpha^*_{j},\tau^*_{j}), \quad 
\alpha^*_{j}=\frac{\alpha_j+\omega\alpha_{3-j}}{\sqrt{1+\alpha_{3-j}^2(1-\omega^2)}}, \quad
\tau_j^{\ast}=\frac{\tau}{\sqrt{1+\alpha_{3-j}^2(1-\omega^2)}},
$$
$$
 X_j|X_{3-j}\sim ESN\left(\omega x_{3-j}, \sqrt{1-\omega^2},\alpha_{j\cdot 3-j}, \tau_{j\cdot 3-j}\right), \quad
\alpha_{j\cdot 3-j}=\alpha_j\sqrt{1-\omega^2}, \quad
\tau_{j\cdot 3-j}=(1-\omega)\alpha_{3-j}x_{3-j}+\tau.
$$
Define $x_j(u)=\Phi^{\leftarrow}(1-u;\alpha^*_{j},\tau^*_{j})$, for any $u\in[0,1]$, where
$\Phi^{\leftarrow}(\cdot;\alpha^*_{j},\tau^*_{j})$ is the inverse  of the marginal distribution function
$\Phi(\cdot;\alpha^*_{j},\tau^*_{j})$, for $j=1,2$. The asymptotic behaviour of $x_j(u)$ as $u\rightarrow 0$ 
is
\begin{equation}\label{eq:quantile}
x_j(u)=\left\{
\begin{tabular}{lc}
$x(u)$, & if\, $\alpha^*_{j}\geq0$\\
$\frac{x(u)}{\bar{\alpha}_{j}}-\frac{\alpha^*_{j} \tau^*_{j}}{\bar{\alpha}_{j}^2}-\frac{\ln(2\sqrt{\pi})+\ln(|\alpha^*_{j}|)+1/2\ln\ln(1/u) +\alpha^{*2}_{j}/2}{\ell_{1/u,\alpha^*_{j}}}$, & if\, $\alpha^*_{j}<0$
\end{tabular}
\right.
\end{equation}
for $j=1,2$, where $\bar{\alpha}_{j}=\{1+\alpha^{*2}_{j}\}^{1/2}$, 
$
x(u) \approx \ell_{1/u,0} -\{\ln(2\sqrt{\pi})+1/2\ln\ln(1/u)+\ln\Phi(\tau_j^*/\bar{\alpha}_{j})\}/\ell_{1/u,0}
$
and $\ell_{1/u,a}=\sqrt{2\ln(1/u)(1+a^2)}$ for any $a\in\real$.
We denote the asymptotic joint survivor function of the bivariate extended skew-normal distribution by
$p(u)=\prob\{X_1>x_1(u),X_2>x_2(u)\}$ for $u\to0$.

For case (i), when $\alpha_1,\alpha_2>0$ then $x_1(u)=x_2(u)=x(u)$.  Set $K=\Phi(\tau/\sqrt{1+\alpha_1^2+\alpha_2^2+2\omega\alpha_1\alpha_2})$. Then, the joint upper tail $p(u)$ behaves as $u\rightarrow 0$ as
\begin{align}\label{eq:joint_tail}
p(u)&=\int_{x(u)}^\infty \left\{1-\Phi\left(\frac{y(u)-\omega v}{\sqrt{1-\omega^2}};\alpha_{1\cdot 2},\tau_{1\cdot 2}\right)\right\}\phi(v;\alpha^*_{2},\tau^*_{2})\der v\nonumber\\
&\approx \frac{\sqrt{1-\omega^2}}{x(u)}\int_0^\infty\frac{\phi(x(u),x(u)+t/x(u);\bar{\bOmega},\balpha,\tau)}
{x(u)(1-\omega)-\omega t/x(u)}\der t\nonumber\\
&\approx \frac{K^{-1}e^{-\frac{x^2(u)}{1+\omega}}}{2\pi(1-\omega)x^2(u)}\left(
\int_0^\infty e^{-\frac{t}{1+\omega}}\der t -\frac{e^{-\frac{x^2(u)(\alpha_1+\alpha_2)^2}{2}-x(u)(\alpha_1+\alpha_2)\tau}}{\sqrt{2\pi}(\alpha_1+\alpha_2)x(u)}
\int_0^\infty e^{-t\{\frac{1}{1+\omega}+\alpha_2(\alpha_1+\alpha_2)\}} \der t
\right)\nonumber\\
&= 
\frac{e^{-x^2(u)/(1+\omega)}(1+\omega)}{2\pi K(1-\omega)x(u)^2}
\left(
1-\frac{e^{-x^2(u)(\alpha_1+\alpha_2)^2/2-x(u)(\alpha_1+\alpha_2)\tau}}{\sqrt{2\pi}(\alpha_1+\alpha_2)\{1+\alpha_2(\alpha_1+\alpha_2)(1+\omega)\}x(u)}
\right).
\end{align}
The first approximation is obtained using Proposition 2.2 from \citet{boris+p+x+s2018}.
The second approximation uses  Mills' ratio approximation.
Substituting $x(u)$ into \eqref{eq:joint_tail} we obtain the approximation
$p(u)\approx u^{1/\eta}\cL(1/u)$ as $u\to 0^+$, where $\eta=(1+\omega)/2$ and
\begin{equation}\label{eq:first_case_sl_fun}
\cL(x)=\frac{(1+\omega)K^{\frac{1-\omega}{1+\omega}}}{(1-\omega)(4\pi\ln 1/u)^{\frac{\omega}{1+\omega}}}
\left(
1-
\frac{(4\pi\ln 1/u)^{\frac{(\alpha_1+\alpha_2)^2-1}{2}}\,u^{(\alpha_1+\alpha_2)^2}K^{(\alpha_1+\alpha_2)^2}e^{-\frac{\tau}{2}}}
{(1+\omega)^{-1}(1-\omega)(\alpha_1+\alpha_2)\{1+\alpha_2(\alpha_1+\alpha_2)(1+\omega)\}}
\right).
\end{equation}
As the second term in the parentheses in \eqref{eq:first_case_sl_fun} is 
$o(u^{(\alpha_1+\alpha_2)^2})$ for $u\to 0^{+}$, then the quantity inside the parentheses $\rightarrow1$ rapidly as $u\rightarrow0^+$, and so $\cL(1/u)$ is well approximated by the first term in 
\eqref{eq:first_case_sl_fun}.
When $\alpha_2<0$ and $\alpha_1\geq-\alpha_2/\omega$, then $\alpha^*_{1},\alpha^*_{2}>0$ and we obtain the same outcome. 

For case (ii), when $\alpha_2<0$ and $-\omega,\alpha_2\leq \alpha_1<-\omega^{-1}\alpha_2$, then $\alpha^*_{1}\geq 0$ and
$\alpha^*_{2}<0$ and hence $x_1(u)=x(u)$ and $x_2(u)$ is given as in the second line of \eqref{eq:quantile}. 
For the case (iia), i.e.~when $\alpha_1>-\bar{\alpha}_{2}\alpha_2$, then following a similar derivation to that of
\eqref{eq:joint_tail}, we obtain that 
$$
p(u)\approx\frac{\bar{\alpha}_{2}^2(1-\omega^2)(1-\omega\bar{\alpha}_{2})^{-1}}
{2\pi^2 K(\bar{\alpha}_{2}-\omega)x^2(u)}
\exp\left[-\frac{x^2(u)}{2}\left\{\frac{1-\omega^2+(\bar{\alpha}_{2}-\omega)^2}{(1-\omega^2)\bar{\alpha}_{2}^2}\right\}\right],
\quad u\rightarrow 0.
$$
Similarly, for the case (iib), i.e.~when $\alpha_1<-\bar{\alpha}_{2}\alpha_2$, by applying
 Mills' ratio we obtain
$$
p(u)\approx\frac{-\bar{\alpha}_{2}^2\{1-\omega\bar{\alpha}_{2} +\alpha_2(\alpha_2+\alpha_1\bar{\alpha}_{2})
(1-\omega^2)\}^{-1}}
{\pi^{3/2}K(\bar{\alpha}_{2}-\omega)(1-\omega^2)^{-1}(\alpha_1+\alpha_2/\bar{\alpha}_{2})x^3(u)}
e^{-\frac{x^2(u)}{2}\left\{\frac{1-\omega^2+(\bar{\alpha}_{2}-\omega)^2}{(1-\omega^2)\bar{\alpha}_{2}^2}
+\left(\alpha_1+\frac{\alpha_2}{\bar{\alpha}_{2}}\right)^2 -\frac{\tau^2}{2}
\right\}},\; u\rightarrow 0.
$$
For case (iii), when $\alpha_2<0$ and $0<\alpha_1<-\omega\alpha_2$, then $\alpha^*_{1},\alpha^*_{2}<0$ and hence 
$x_1(u)$ and $x_2(u)$ are given as in the second line of \eqref{eq:quantile}. Then, by Proposition 2.2 from \citet{boris+p+x+s2018}
we obtain 
\begin{align*}
p(u)&\approx\frac{-\bar{\alpha}_{2}^{3/2}\bar{\alpha}_{1}^2(1-\omega^2)
(\bar{\alpha}_{2}-\omega\bar{\alpha}_{1})^{-1}(\alpha_1\bar{\alpha}_{2}+\alpha_2\bar{\alpha}_{1})^{-1}}
{(2\pi)^{3/2}K\{1-\omega\bar{\alpha}_{2}+\alpha_2(\alpha_2+\alpha_1\bar{\alpha}_{2}/\bar{\alpha}_{1})(1-\omega^2)\}x^3(u)}\\
&\times
\exp\left[-\frac{x^2(u)}{2(1-\omega^2)}
\left(
\frac{\alpha_1^2(1-\omega^2)+1}{\bar{\alpha}_{1}^2}+
\frac{\alpha_2^2(1-\omega^2)+1}{\bar{\alpha}_{2}^2}+
\frac{2(\alpha_1\alpha_2(1-\omega^2)-\omega)}{\bar{\alpha}_{1}\bar{\alpha}_{2}}
\right)
-\frac{\tau^2}{2}
\right]
\quad u\rightarrow 0.
\end{align*}
When $\alpha_1,\alpha_2<0$ and $\omega_2^{-1}\alpha_2\leq \alpha_1<0$  the same argument holds.
Finally, interchanging $\alpha_1$ with $\alpha_2$ produces the same results but 
where $\alpha_{j}$ and $\bar{\alpha}_{j}$ are substituted in the above with $\alpha_{3-j}$ and $\bar{\alpha}_{3-j}$ respectively, for $j=1,2.$

\subsection{Proof of Theorem~\ref{theo:tri_array}}

Let $\bX_{n,m}\sim ESN_d(\bar{\bOmega}_n,\balpha_n,\tau)$, $n\in\nat$ and $m=1,\ldots,n$, where $\bar{\bOmega}_n$ and $\balpha_n$ are defined in Condition~\ref{ass:tri_array} and $\tau\in\real$. 
We want to derive norming constants $\bba_n>\bzero$ and $\bb_n\in\real^d$ such that we can derive a non-trivial limit distribution for $\Phi_d^n(\bba_n\bx+\bb_n;\bar{\bOmega}_n,\balpha_n,\tau)$. 
Recall that from \citet{arellano2010} we have that for all $j\in I$, $X_{n,m;j}\sim ESN(\alpha_{n;j}^{*},\tau_{n;j}^{*})$, where
$\alpha_{n;j}^{*}$ and $\tau_{n;j}^{*}$ are appropriate slant and extension marginal parameters.
Then, we may state the following result.
\begin{lem}\label{lem:first_step}
For all $j\in I$ define the normalising constants $a_{n;j} = \ell_{n}^{-1}$, 
\begin{align*}
b_{n;j} &= \ell_{n}-\frac{\ln(2\sqrt{\pi})+(1/2)\ln\ln n  +\ln\Phi\left(\tau_{n;j}^{\ast}/\bar{\alpha}_{n;j}\right)-\ln\Phi\left(\alpha_{n;j}^{\ast}\ell_{n}+\tau_{n;j}^{\ast}\right)}
{\ell_{n}},\qquad \qquad \quad\mbox{if } \alpha^{\ast}_{n;j}\geq 0,\\
b_{n;j} &= \ell_{n}
-\frac{\ln \sqrt{2\pi}+\ln\Phi\left(\tau_{n;j}^{\ast}/\bar{\alpha}_{n;j}\right)-\ln\Phi\left(\alpha_{n;j}^{\ast}\ell_{n}+\tau_{n;j}^{\ast}\right)}
{\ell_{n}}-\frac{\ln\Phi\left(\bar{\alpha}_{n;j}^2\ell_{n}+\alpha_{n;j}^{\ast}\tau_{n;j}^{\ast}\right)}{\ell_{n}},\;  \mbox{if }\alpha_{n;j}^{\ast} < 0,
\end{align*}
where $\bar{\alpha}_{n;j}=\{1 + \alpha_{n;j}^{\ast 2}\}^{1/2}$, $\ell_{n}=\sqrt{2\ln n}$. Then, for all $j\in I$,
$$
\lim_{n\to\infty} \Phi^n_{\alpha_{n;j}^{*},\tau_{n;j}^{*}}(a_{n;j}x_j+b_{n;j})= e^{-e^{-x_j}}, \quad x_j\in\real.
$$
\end{lem}
Since for all $j\in I$, $e^{-e^{-x_j}}$ is continuous then the weak convergence of $ESN_d(\bar{\bOmega}_n,\balpha_n,\tau)$ is equivalent to weak convergence of the marginal distributions functions
and the copula function \citep[e.g.][Section~8.3.2]{beirlant2004}. It remains to derive the limiting form of the copula function of $ESN_d(\bar{\bOmega}_n,\balpha_n,\tau)$. We complete the proof deriving the stable-tail dependence function $L$, since  an extreme-value copula
is of the form $C(\bu)=\exp\{-L(-\ln u_1,\ldots,-\ln u_d)\}$ (see Section~\ref{sec:back}).
\begin{lem}\label{lem:second_step}
The stable-tail dependence function associated with  the limit distribution of $\Phi_d^n(\bba_n\bx+\bb_n;\bar{\bOmega}_n,\balpha_n,\tau)$ is
\begin{equation*}\label{eq:stable-tail_1}
\begin{split}
L(\bz)&=\lim_{n\to\infty}n\left\{1-\Pr\left(\Phi_{\alpha_{n;j}^{*},\tau_{n;j}^{*}}(X_j)\leq 1-\frac{z_j}{n},j=1,\ldots,d\right)\right\},\quad \bz\in[0,\infty)^d\\
&=\sum_{j=1}^d z_j \, \Phi_{d-1}
\left\{
\left(
\lambda_{ij} + \frac{1}{2 \lambda_{ij}} \log \frac{\tilde{z}_j}{\tilde{z}_i},i \in I_j\right)^{\top}; 
\bar{\bLambda}_j,\tilde{\balpha}_j,\tilde{\tau}_j
\right\},
\end{split}
\end{equation*}
where for all $j\in I$, $\bar{\bLambda}_j$, $\tilde{\balpha}_j$ and $\tilde{\tau}_j$ are given in statement of the theorem.
\end{lem}
\bibliographystyle{chicago}
\bibliography{biblio}

\end{document}


\title{Supplementary material for `Extremal properties of the multivariate extended skew-normal distribution'}

\author{B. Beranger\footnote{School of Mathematics and Statistics, University of New South Wales, Sydney, Australia.}\;\,\footnote{Communicating Author: {\tt B.Beranger@unsw.edu.au}},\; S. A. Padoan\footnote{Department of Decision Sciences, Bocconi University of Milan, Italy.
},\; Y. Xu$^*$\; and\; S. A. Sisson$^*$
}

\date{}

\maketitle
\begin{abstract}
This document contains the proof of Lemma 1 used in proof of Proposition 3.1, and Lemmas 2 and 3 used in the
proof of Theorem 3.1 in the main paper.
\end{abstract}

Note that all references below of the form ($\cdot^*$) refer to equation ($\cdot$) in the main paper.
When we refer to a proposition or theorem we implicitly refer to a result in the paper unless otherwise specified. 
\section{Technical details}\label{sec:details}

We first recall  properties  of the extended
skew-normal distribution that will be useful in some the following proofs \citep[e.g.][]{arellano2010}.
\begin{proper}\label{lem:prop_esn}
Let $\bX\sim ESN_d(\bmu, \bOmega, \balpha, \tau)$.
Let $I \subset \{ 1, \ldots, d\}$ and $\bar{I} = \{ 1, \ldots, d\} \backslash I$ 
identify the $d_I$ and $d_{\bar{I}}$-dimensional subvector partition of $\bX$ such that 
$\bX = \left(\bX_I^\top, \bX_{\bar{I}}^\top \right)^\top$, 
with corresponding partition of the parameters $(\bmu, \bOmega, \balpha)$. Then
\begin{inparaenum}
\item \label{en:second_cond_mar} $\bX_I \sim ESN_{d_I}(\bmu_I, \bOmega_{II}, \balpha_I^*, \tau_I^*)$ where
$
\balpha_I^* = \frac{\balpha_I + \bar{\bOmega}_{II}^{-1} \bOmega_{I \bar{I}} \balpha_{\bar{I}} }
{\sqrt{1 + Q_{\tilde{\sbOmega}_{\bar{I}\bar{I}\cdot I} (\balpha_{\bar{I}}) } }}
$
and 
$
\tau_I^* = \frac{\tau }
{\sqrt{1 + Q_{\tilde{\sbOmega}_{\bar{I}\bar{I}\cdot I} (\balpha_{\bar{I}}) } }}
$,
given 
$\tilde{\bOmega}_{\bar{I}\bar{I}\cdot I} = 
\bar{\bOmega}_{\bar{I}\bar{I}} - 
\bar{\bOmega}_{\bar{I}I} \bar{\bOmega}_{II}^{-1} \bar{\bOmega}_{I\bar{I}}$
and $Q_{\tilde{\sbOmega}_{\bar{I}\bar{I}\cdot I} (\balpha_{\bar{I}}) } = 
\balpha_{\bar{I}}^\top \tilde{\sbOmega}_{\bar{I}\bar{I}\cdot I}^{-1} \balpha_{\bar{I}}$. 
\item \label{en:second_cond_cond} $(\bX_{\bar{I}} | \bX_I = \bx_I) 
\sim ESN_{d_{\bar{I}}}(\bmu_{\bar{I} \cdot I}, \bOmega_{\bar{I} \cdot I}, 
\balpha_{\bar{I} \cdot I}, 
\tau_{\bar{I} \cdot I})$ where
$\bmu_{\bar{I} \cdot I} = \bmu_{\bar{I}} + 
\bar{\bOmega}_{\bar{I}I} \bar{\bOmega}_{II}^{-1} (\bx_I - \bmu_I)$,
$\bOmega_{\bar{I} \cdot I} = \bOmega_{\bar{I}\bar{I}} - 
\bOmega_{\bar{I}I} \bOmega_{II}^{-1} \bOmega_{I\bar{I}}$,
$\balpha_{\bar{I} \cdot I} = \bomega_{\bar{I} \cdot I} \bomega_{\bar{I}}^{-1} \balpha_{\bar{I}}$,
$\bomega_{\bar{I} \cdot I} = \diag (\bOmega_{\bar{I} \cdot I})^{1/2}$,
$\bomega_{\bar{I}} = \diag (\bOmega_{\bar{I} \bar{I}})^{1/2}$
and
$\tau_{\bar{I} \cdot I} = \left( \balpha_{\bar{I}}^\top \bar{\bOmega}_{\bar{I}I} \bar{\bOmega}_{II}^{-1} + \balpha_I^\top \right) \bx_I + \tau$.
\end{inparaenum}
\end{proper}

%
%
%
\subsection{Proof of Lemma 1}
Recall that if $\bX\sim ESN_d(\bar{\bOmega},\balpha,\tau)$, then from Property \ref{lem:prop_esn}\ref{en:second_cond_mar}
and  \ref{lem:prop_esn}\ref{en:second_cond_cond}
we have that for any pair $(X_i,X_j)$ with $1\leq i<j\leq d$,
\begin{eqnarray*}
X_i&\sim ESN(\alpha_i^{\ast}, \tau_i^{\ast}),\quad 
\alpha_i^{\ast}=\frac{\alpha_i+\omega_{i,j}\alpha_j}{\sqrt{1+\alpha_j^2(1-\omega^2_{i,j})}},\quad
\frac{\tau}{\sqrt{1+\alpha_j^2(1-\omega_{i,j}^2)}}\\
X_j&\sim ESN(\alpha_j^{\ast}, \tau_j^{\ast}),\quad 
\alpha_j^{\ast}=\frac{\alpha_j+\omega_{i,j}\alpha_i}{\sqrt{1+\alpha_i^2(1-\omega^2_{i,j})}},\quad
\frac{\tau}{\sqrt{1+\alpha_i^2(1-\omega_{i,j}^2)}}.
\end{eqnarray*}
We consider the scenarios: (a) $0<\alpha_i^{\ast}<\alpha_j^{\ast}$,  (b)
$\alpha_i,\alpha_j<0$ and assume  $\alpha_i^{\ast}<\alpha_j^{\ast}$ with
$\alpha_i^{\ast}<0$, (c) $\alpha_i^{\ast},\alpha_j^{\ast}<0$ and assume that $\alpha_i<0$ and
$\alpha_j\geq 0$
and (d) $\alpha_i^{\ast}<0$ and $\alpha_j^{\ast}\geq 0$. For cases (a) and (b) we need to show that
\begin{equation}\label{eq:asym_ind_ineq}
\lim_{x\to\infty}\frac{\Pr(X_j\geq x, X_i\geq x)}{\Pr(X_i\geq x)}=0.
\end{equation}
With case (a), we know that $\Pr(X_j\geq x, X_i\geq x)\leq \Pr(\min(Z_i,Z_j)\geq x)/K$, 
where 
\begin{equation}\label{eq_details}
(Z_i,Z_j)\sim N_2(\bar{\bOmega}), \quad K=\Phi(\tau/\sqrt{1+\alpha_i^2+\alpha_j^2+2\omega_{i,j}\alpha_i\alpha_j}).
\end{equation}
Furthermore, 
$$
\min(Z_i,Z_j)\sim ESN(\alpha^{\ast},0), \quad \alpha^{\ast}=-\sqrt{(1-\omega_{i,j})/(1+\omega_{i,j})},
$$
see e.g.~\citet[][p. 29]{azzalini2014}. Therefore by applying these results to \eqref{eq:asym_ind_ineq} and applying
Proposition~2.2 from \citet{boris+p+x+s2018} to $1-\Phi_{\alpha_i^{\ast},\tau_i^{\ast}}(x)$ 
and $1-\Phi_{\alpha^{\ast},0}(x)$ we obtain
\begin{equation*}
\begin{split}
\frac{\Pr(X_j\geq x, X_i\geq x)}{\Pr(X_i\geq x)}&\leq\lim_{x\to\infty}\frac{1-\Phi_{\alpha^{\ast},0}(x)}
{1-\Phi_{\alpha_i^{\ast},\tau_i^{\ast}}(x)}K^{-1}\\
&\leq\lim_{x\to\infty}\frac{(1-\omega_{i,j})^{3/2}\phi\left(x\sqrt{2/(1+\omega_{i,j})}\right)}{\sqrt{2\pi(1-\omega_{i,j})}x}\frac{\Phi\left(\tau_i^{\ast}/\bar{\alpha}_i\right)}{\phi(x)\Phi(\alpha_i^{\ast}x+\tau_i^{\ast})}K^{-1}\\
&=0
\end{split}
\end{equation*}
and therefore the limit in \eqref{eq:asym_ind_ineq} is satisfied.

With case (b), using similar arguments than above we have
\begin{eqnarray*}
\Pr(X_j\geq x, X_i\geq x) &\leq&  \Pr(Z_j\geq x, Z_i\geq x)\Phi(x(\alpha_i+\alpha_j)+\tau)K^{-1}\\ 
&=&(1-\Phi_{\alpha^{\ast},0}(x))\Phi(x(\alpha_i+\alpha_j)+\tau) K^{-1}.
\end{eqnarray*}
Since $x(\alpha_i+\alpha_j)+\tau <0$ as $x\to\infty$, then by Proposition 2.2 from \citet{boris+p+x+s2018} we have
$$
\Phi(x(\alpha_i+\alpha_j)+\tau)\approx-\phi(x(\alpha_i+\alpha_j)+\tau)/(x(\alpha_i+\alpha_j)+\tau), \quad x\to\infty.
$$
Furthermore, since $\alpha_i^{\ast}<0$ then $\alpha_i^{\ast}x+\tau_i^{\ast}<0$
as $x\to\infty$ and by Proposition 2.2 from \citet{boris+p+x+s2018} we have
$$\Phi(\alpha_i^{\ast}x+\tau_i^{\ast})=1-\Phi(-\alpha_i^{\ast}x-\tau_i^{\ast})\approx-\phi(\alpha_i^{\ast}x+\tau_i^{\ast})/(\alpha_i^{\ast}x+\tau_i^{\ast}),\quad x\to\infty.
$$
These results imply that
\begin{equation*}
\begin{split}
\frac{\Pr(X_j\geq x, X_i\geq x)}{\Pr(X_i\geq x)}&\leq\lim_{x\to\infty}\frac{(1-\omega_{i,j})^{3/2}\phi\left(x\sqrt{2/(1+\omega_{i,j})}\right)}{\sqrt{2\pi(1-\omega_{i,j})}x}\frac{\phi\left(x(\alpha_i+\alpha_j)+\tau\right)}{\phi(x)\phi(\alpha_i^{\ast}x+\tau_i^{\ast})}\frac{\alpha_i^{\ast}x+\tau_i^{\ast}}{x(\alpha_i+\alpha_j)+\tau}\\
&\approx \lim_{x\to\infty}\frac{-(1-\omega_{i,j})^{3/2}\alpha_i^{\ast}}{\sqrt{2\pi(1-\omega_{i,j})}x}\frac{\phi\left(x\sqrt{2/(1+\omega_{i,j})}\right)\phi\left(x(\alpha_i+\alpha_j)+\tau\right)}{\phi(x)\phi(\alpha_i^{\ast}x+\tau_i^{\ast})}\\
&=0,
\end{split}
\end{equation*}
where in the last step used the fact that
\begin{equation*}
\begin{split}
\frac{\phi\left(x\sqrt{2/(1+\omega_{i,j})}\right)\phi\left(x(\alpha_i+\alpha_j)+\tau\right)}{\phi(x)\phi(\alpha_i^{\ast}x+\tau_i^{\ast})}&=
\exp\left(
-\frac{1}{2}\left(
\frac{2x^2}{1+\omega_{i,j}}+(x(\alpha_i+\alpha_j)+\tau)^2-x^2-(\alpha_i^{\ast}x+\tau_i^{\ast})^2
\right)
\right)\\
&\approx\exp\left[
-\frac{1}{2}\left\{x^2
\left(
\frac{2}{1+\omega_{i,j}}+(\alpha_i+\alpha_j)^2-1-\alpha_i^{\ast 2}
\right)
\right\}
\right],\;x\to\infty,
\end{split}
\end{equation*}
and where $2/(1+\omega_{i,j})+(\alpha_1+\alpha_2)^2-1-\alpha_i^{\ast 2}>0$. Therefore, also in this case the limit in \eqref{eq:asym_ind_ineq} is satisfied.

With case (c) we need to show that
\begin{equation}\label{eq:chi_asym}
\lim_{x\to\infty}\frac{\Pr\left(X_j\geq x, X_i\geq x\bar{\alpha}_j/\bar{\alpha}_i\right)}
{\Pr\left(X_i\geq x\bar{\alpha}_j/\bar{\alpha}_i\right)}=0.
\end{equation}
Applying Proposition 2.2 from \citet{boris+p+x+s2018} to $\Phi(\alpha_i^{\ast}\bar{\alpha}_j/\bar{\alpha}_ix+\tau_i^{\ast})$
we obtain for the denominator that as $x\to\infty$
\begin{equation*}
\begin{split}
D_1(x)=\frac{\partial}{\partial x}
\Pr\left(X_i\geq x\bar{\alpha}_j/\bar{\alpha}_i\right)&=
-\bar{\alpha}_j/\bar{\alpha}_i
\phi_{\alpha_i^{\ast},\tau_i^{\ast}}\left(
x\bar{\alpha}_j/\bar{\alpha}_i
\right)\\
&\approx
\bar{\alpha}_j/\bar{\alpha}_i
\frac{\phi\left(
x\bar{\alpha}_j/\bar{\alpha}_i
\right)\phi\left(
\alpha_i^{\ast}\bar{\alpha}_j/\bar{\alpha}_ix+\tau_i^{\ast}
\right)}{\Phi(\tau_i^{\ast}/\bar{\alpha}_i)\left(\alpha_i^{\ast}\bar{\alpha}_j/\bar{\alpha}_ix+\tau_i^{\ast}\right)}\\
&\approx \bar{\alpha}_j/\bar{\alpha}_i
\frac{\bar{\alpha}_i\exp\left\{-x^2\bar{\alpha}_j^2/2\right\}}
{2\pi x\alpha_i^{\ast}\bar{\alpha}_j\Phi(\tau_i^{\ast}/\bar{\alpha}_i)
}.
\end{split}
\end{equation*}
For the numerator we have
\begin{equation*}
\begin{split}
\frac{\partial}{\partial x}
\Pr\left(X_j\geq x, X_i\geq x\bar{\alpha}_j/\bar{\alpha}_i\right)&=
-\bar{\alpha}_j/\bar{\alpha}_i\int_x^{\infty}
\phi\left(\left(x\bar{\alpha}_j/\bar{\alpha}_i, v\right)^{\top};
\bar{\bOmega},\balpha,\tau\right)\diff v
-\int_{a}^{\infty}
\phi\left((x,v)^{\top};\bar{\bOmega},\balpha,\tau\right)\diff v\\ 
&=D_2(x)+D_3(x),
\end{split}
\end{equation*}
where $a=x\bar{\alpha}_j/\bar{\alpha}_i$.
For $D_2(x)$ we use integration by parts where
$$
r=-\frac{\Phi\left(\alpha_i x\bar{\alpha}_j/\bar{\alpha}_i 
+\alpha_j v+\tau \right)}{v+\omega_{i,j} x\bar{\alpha}_j/\bar{\alpha}_i}\quad
s=\phi\left(\left(x\bar{\alpha}_j/\bar{\alpha}_i, v\right)^{\top};
\bar{\bOmega}\right),
$$
and then applying Proposition 2.2 from \citet{boris+p+x+s2018} to 
$\Phi\left(\alpha_i x\bar{\alpha}_j/\bar{\alpha}_i 
+\alpha_j v+\tau \right)$ we obtain
$$
D_2(x)\approx \frac{\bar{\alpha}_j}{\bar{\alpha}_i}\,
\frac{\phi\left(\left(x\bar{\alpha}_j/\bar{\alpha}_i, x\right)^{\top};
\bar{\bOmega}\right)\phi\left(\alpha_i x\bar{\alpha}_j/\bar{\alpha}_i 
+\alpha_j x+\tau \right)}
{\left(x+\omega_{i,j} x\bar{\alpha}_j/\bar{\alpha}_i\right)
\left(
\alpha_i x\bar{\alpha}_j/\bar{\alpha}_i 
+\alpha_j x+\tau
\right)}K^{-1}\quad\mbox{as } x\to\infty.
$$
For $D_3(x)$ we also use integration by parts where
$$
r=-\frac{\Phi(\alpha_1 v +\alpha_2 x +\tau)}{v+\omega_{i,j}x},\quad s=\phi\left(\left(x, v\right)^{\top};
\bar{\bOmega}\right),
$$
and then applying Proposition 2.2 from \citet{boris+p+x+s2018} to 
$\Phi\left(\alpha_i x\bar{\alpha}_j/\bar{\alpha}_i 
+\alpha_j v+\tau \right)$ we obtain
$$
D_3(x)\approx
\frac{\phi\left(\left(x\bar{\alpha}_j/\bar{\alpha}_i, x\right)^{\top};
\bar{\bOmega}\right)\phi\left(\alpha_i x\bar{\alpha}_j/\bar{\alpha}_i 
+\alpha_j x+\tau \right)}
{\left(\omega_{i,j}x+ x\bar{\alpha}_j/\bar{\alpha}_i\right)
\left(
\alpha_i x\bar{\alpha}_j/\bar{\alpha}_i 
+\alpha_j x+\tau
\right)}K^{-1},\quad\mbox{as } x\to\infty.
$$
Then the ratio in \eqref{eq:chi_asym} behaves asymptotically as $(D_2(x)+D_3(x))/D_1(x)$
as $x\to \infty$. Furthermore, as $x\to\infty$ 
\begin{align*}
\frac{D_2(x)}{D_1(x)}&\approx
\frac{\phi\left(\left(x\bar{\alpha}_j/\bar{\alpha}_i, x\right)^{\top};
\bar{\bOmega}\right)\phi\left(\alpha_i x\bar{\alpha}_j/\bar{\alpha}_i 
+\alpha_j x+\tau \right)}
{x\left(1+\omega_{i,j} \bar{\alpha}_j/\bar{\alpha}_i\right)
\left(
\alpha_i \bar{\alpha}_j/\bar{\alpha}_i 
+\alpha_j +\tau/x
\right)\exp\left(-x^2\bar{\alpha}_j^2/2\right)}\\ 
&\to 0, \quad x\to\infty,
\end{align*}
and as $x\to\infty$
\begin{align*}
\frac{D_3(x)}{D_1(x)}&\approx
\frac{\phi\left(\left(x\bar{\alpha}_j/\bar{\alpha}_i, x\right)^{\top};
\bar{\bOmega}\right)\phi\left(\alpha_i x\bar{\alpha}_j/\bar{\alpha}_i 
+\alpha_j x+\tau \right)}
{x\left(\omega_{i,j} +\bar{\alpha}_j/\bar{\alpha}_i\right)
\left(
\alpha_i \bar{\alpha}_j/\bar{\alpha}_i 
+\alpha_j +\tau/x
\right)\exp\left(-x^2\bar{\alpha}_j^2/2\right)}\\ 
&\to 0.
\end{align*}
Therefore, the limit in \eqref{eq:chi_asym} is proven.

With case (d) we need to show that
 \begin{equation}\label{eq:last_limit}
 \lim_{x\to\infty}\frac{\Pr\left(X_j\geq x, X_i\geq x/\bar{\alpha}_i\right)}
{\Pr\left(X_i\geq x/\bar{\alpha}_i\right)}=0.
 \end{equation}
 Note that in this case as $x\to\infty$ we have 
$$
\Phi\left(\alpha_j^{\ast}\bar{\alpha}_ix+\tau^{\ast}\right)>
\frac{1}{\sqrt{2\pi}(-\alpha_i^{\ast})\bar{\alpha}_jx},
$$
which implies that as $x\to\infty$,
$$
1-\Phi_{x \alpha_j^{\ast},\tau_{j}^{\ast}}\left(\bar{\alpha}_i\right)>
1-\Phi_{\alpha_i^{\ast},\tau_{i}^{\ast}}(x),\quad
\Phi_{\alpha_j^{\ast},\tau_{j}^{\ast}}\left(x \bar{\alpha}_i\right)<
\Phi_{\alpha_i^{\ast},\tau_{i}^{\ast}}(x).
$$
These results imply  that
\begin{align*}
\frac{\Pr\left(X_j\geq x, X_i\geq x/\bar{\alpha}_i\right)}
{\Pr\left(X_i\geq x/\bar{\alpha}_i\right)}
\leq
\frac{\Pr\left(Z_i\geq x, Z_j\geq x/\bar{\alpha}_i\right)}
{\Pr\left(X_i\geq x/\bar{\alpha}_i\right)}K^{-1}
\end{align*}
where $(Z_i,Z_j)$ and $K$ are given in \eqref{eq_details}. By Savage's approximation and applying
Proposition 2.2 from \citet{boris+p+x+s2018} we obtain as $x\to\infty$
\begin{align*}
\frac{\Pr\left(Z_i\geq x, Z_j\geq x/\bar{\alpha}_i\right)}
{\Pr\left(X_i\geq x/\bar{\alpha}_i\right)}K^{-1}
&\leq  \frac{\phi((x,x/\bar{\alpha}_i)^{\top};\bar{\bOmega})}
{x^2\left(1/\bar{\alpha}_i-\omega_{i,j}\right)\left(1-\omega_{i,j}/\bar{\alpha}_i\right)\Pr\left(X_i\geq x/\bar{\alpha}_i\right)
}K^{-1}\\
&\approx \frac{\sqrt{2\pi}(-\alpha_i^{\ast})}{\left(1/\bar{\alpha}_i-\omega_{i,j}\right)\left(1-\omega_{i,j}/\bar{\alpha}_i\right)}\frac{\phi((x,x/\bar{\alpha}_i)^{\top};\bar{\bOmega})}{\phi(x)}\\
&\approx \frac{(1+\alpha_i^{\ast^{2}})(-\alpha_i^{\ast})}
{
\left(1/\bar{\alpha}_i-\omega_{i,j}\right)\left(1-\omega_{i,j}/\bar{\alpha}_i\right)(1-\omega_{i,j}^2)}\exp\left\{-
\frac{x^2}{2}\frac{\left(1+\omega_{i,j}\bar{\alpha}_i\right)^2}{(1-\omega_{i,j}^2)(1+\alpha_i^{\ast^{2}})}
\right\}\\
&\to 0,\quad x\to\infty,
\end{align*}
and therefore also the last limit in \eqref{eq:last_limit} is proven.

%

\subsection{Proof of Lemma~2}

Let $\bX_{n,m}\sim ESN_d(\bar{\bOmega}_n,\balpha_n,\tau)$, $n\in\nat$ with $m=1,\ldots,n$, where $\bar{\bOmega}_n$ and $\balpha_n$ are defined in Condition~1 and $\tau\in\real$. From Property \ref{lem:prop_esn}\ref{en:second_cond_mar} we have
\begin{equation}\label{eq_marginals}
X_{n,m;j}\sim ESN(\alpha_{n;j}^{*},\tau_{n;j}^{*}),\quad j\in I,
\end{equation}
where $I=\{1,\ldots,d\}$,
\begin{equation}\label{eq:marg_par}
\alpha_{n;j}^*=\frac{\alpha_{n;j}+\sum_{i\in I_j}\alpha_{n;i}\omega_{n;i,j}}{C(\bar{\bOmega}_n,\balpha_n)},\quad
\tau_{n,j}^*=\frac{\tau}{C(\bar{\bOmega}_n,\balpha_n)},
\end{equation}
with
$$
C(\bar{\bOmega}_n,\balpha_n)=\left\{1+\sum_{i,k\in I_j}
\alpha_{n;i}\alpha_{n;k}(\omega_{n;i,k}-\omega_{n;i,j}\omega_{n;j,k})\right\}^{1/2}
$$
and
\begin{equation}\label{eq_conditionals}
(X_{n,m;i},i \in I_j)^{\top}|X_{n,m;j}=x_{j}\sim ESN_{d-1}(\tilde{\bmu}_{n;j},\tilde{\bOmega}_{n;j},\tilde{\balpha}_{n;j},\tilde{\tau}_{n;j}),\quad j\in I,
\end{equation}
where $\tilde{\bmu}_{n;j}=(x_j\omega_{n;i,j},i\in I_j)^{\top}$, $I_j=I\backslash\{j\}$, 
$\tilde{\bOmega}_{n;j}$ is a $(d-1)\times(d-1)$ correlation matrix
with diagonal entries $1-\omega_{n;i,j}^2$ for $i \in I_j$ and upper diagonal entries $\omega_{n;i,k}-\omega_{n;i,j}\omega_{n;j,k}$ for $i,k\in I_j$ and
\begin{equation}\label{eq:cond_par}
\tilde{\balpha}_{n;j}=\left(\left(1-\omega_{n;i,j}^2\right)^{1/2}\alpha_{n;i}, i \in I_j\right)^{\top},\quad 
\tilde{\tau}_{n;j}=\left(\sum_{i\in I_j}(\alpha_{n;i}\omega_{n;i,j})+\alpha_{n;j}\right)x_j+\tau.
\end{equation}

Using similar steps to those in the proof of Proposition 2.4 from \citet{boris+p+x+s2018} we may derive
for all $j \in I $ the normalising constants $a_{n;j} = \ell_{n}^{-1}$, 
\begin{align*}
b_{n;j} &= \ell_{n}-\frac{\ln(2\sqrt{\pi})+(1/2)\ln\ln n  +\ln\Phi\left(\tau_{n;j}^{\ast}/\bar{\alpha}_{n;j}\right)-\ln\Phi\left(\alpha_{n;j}^{\ast}\ell_{n}+\tau_{n;j}^{\ast}\right)}
{\ell_{n}},\qquad \qquad \quad\mbox{if } \alpha^{\ast}_{n;j}\geq 0,\\
b_{n;j} &= \ell_{n}
-\frac{\ln \sqrt{2\pi}+\ln\Phi\left(\tau_{n;j}^{\ast}/\bar{\alpha}_{n;j}\right)-\ln\Phi\left(\alpha_{n;j}^{\ast}\ell_{n}+\tau_{n;j}^{\ast}\right)}
{\ell_{n}}-\frac{\ln\Phi\left(\bar{\alpha}_{n;j}^2\ell_{n}+\alpha_{n;j}^{\ast}\tau_{n;j}^{\ast}\right)}{\ell_{n}},\;  \mbox{if }\alpha_{n;j}^{\ast} < 0,
\end{align*}
where $\bar{\alpha}_{n;j}=\{1 + \alpha_{n;j}^{\ast 2}\}^{1/2}$, $\ell_{n}=\sqrt{2\ln n}$. Once again with similar arguments to those in the proof of Proposition 2.4 from \citet{boris+p+x+s2018} we obtain for all $j\in I$,
$$
\lim_{n\to\infty} \Phi^n_{\alpha_{n;j}^{*},\tau_{n;j}^{*}}(a_{n;j}x_j+b_{n;j})= e^{-e^{-x_j}}, \quad x_j\in\real.
$$

\subsection{Proof of Lemma~3}

Since the second-order partial derivatives of the copula function of $ESN_d(\bar{\bOmega}_n,\balpha_n,\tau)$ are continuous, then
from the theory of  multivariate tail dependence functions \citep[e.g.][]{Nikolou+joe09,Li09} 
the stable-tail dependence function can be derived as 
\begin{equation*}\label{eq:stable-tail_1}
\begin{split}
L(\bz)&=\lim_{n\to\infty}n\left\{1-\Pr\left(\Phi_{\alpha_{n;j}^{*},\tau_{n;j}^{*}}(X_j)\leq 1-\frac{z_j}{n},j=1,\ldots,d\right)\right\},\quad \bz\in[0,\infty)^d\\
&
=\sum_{j=1}^d z_j\lim_{n\to\infty}
\Pr\left(\Phi_{\alpha_{n;i}^{*},\tau_{n;i}^{*}}(X_i)\leq 1-\frac{z_i}{n},i\in I_j|\Phi_{\alpha_{n;j}^{*},\tau_{n;j}^{*}}(X_j)=1-\frac{z_j}{n}\right)\\
&=\sum_{j=1}^d z_j\lim_{n\to\infty}
\Pr\left(X_i\leq a_{n;i}\left(\frac{U_i\left(\frac{n}{z_i}\right)-b_{n;i}}{a_{n;i}}\right)+b_{n;i},i\in I_j|X_j=a_{n;j}\left(\frac{U_j\left(\frac{n}{z_j}\right)-b_{n;j}}{a_{n;j}}\right)+b_{n;j}\right)
\end{split}
\end{equation*}
where 
$$
U_i(n/z_i)=\Phi_{\alpha_{n;i}^{*},\tau_{n;i}^{*}}^{\leftarrow}(1-z_i/n), \quad
U_j(n/z_j)=\Phi_{\alpha_{n;j}^{*},\tau_{n;j}^{*}}^{\leftarrow}(1-z_j/n)
$$ 
and $\Phi_{\alpha_{n;i}^{*},\tau_{n;i}^{*}}^{\leftarrow}$,$\Phi_{\alpha_{n;j}^{*},\tau_{n;j}^{*}}^{\leftarrow}$ denote the left-continuous inverse of $\Phi_{\alpha_{n;i}^{*},\tau_{n;i}^{*}}$ 
and $\Phi_{\alpha_{n;j}^{*},\tau_{n;j}^{*}}$, for all $j\in I$ and $i\in I_j$. 

From Condition~1 we obtain the following results. For all $j\in I$ and $i,k\in I_j$, as $n\to\infty$ we have 
\begin{align*}
C(\bar{\bOmega}_n,\balpha_n)\to 
\left\{1+\sum_{i,k\in I_j}
\alpha_{i}^{\circ}\alpha_{k}^{\circ}(\lambda_{i,j}^2+\lambda_{j,k}^2-\lambda^2_{i,k})\right\}^{1/2}=:C(\bar{\bLambda},\balpha^{\circ}),
\end{align*}
where $\alpha_{i}^{\circ}\alpha_{k}\in\real$, $\lambda_{i,j}^2, \lambda_{j,k}^2,\lambda^2_{i,k}\in(0,\infty]$. Furthermore, as $n\to\infty$,
\begin{align}
\label{eq:mat_conv}\frac{\omega_{n;i,k}-\omega_{n;i,j}\omega_{n;j,k}}
{\sqrt{(1-\omega_{n;i,j}^2)(1-\omega_{n;k,j}^2)}}
&\approx\frac{1-\frac{\lambda_{i,k}^2}{\ln n}-\left(1-\frac{\lambda_{i,j}^2}{\ln n}\right)\left(1-\frac{\lambda_{j,k}^2}{\ln n}\right)}
{\sqrt{\left\{1-\left(1-\frac{\lambda_{i,j}^2}{\ln n}\right)^2\right\}\left\{1-\left(1-\frac{\lambda_{j,k}^2}{\ln n}\right)^2\right\}}}\\ 
&\nonumber\to
\frac{-\lambda_{i,k}^2+\lambda_{i,j}^2+\lambda_{j,k}^2}{2\lambda_{i,j}\lambda_{j,k}},
\end{align}
\begin{align}
\label{eq:alpha_conv}\sqrt{1-\omega_{n;i,j}^2}\alpha_{n;i}&=
\frac{\sqrt{(1+\omega_{n;i,j})(1-\omega_{n;i,j})\ln n}\alpha_{n;i}}{\ln n}\\
&\nonumber\to \sqrt{2}\lambda_{i,j}\alpha_i^{\circ},
\end{align}
\begin{align}
\label{eq:marg_alpha_conv}\alpha_{n;j}^{\ast}=
\frac{\sum_{j\in I}\alpha_{n;j}-\sum_{i\in I_j}\alpha_{n;i}(1-\omega_{n;i,j})}{C(\bar{\bOmega}_n,\balpha_n)}
\to 0,
\end{align}
\begin{align}
\label{eq:marg_ext_conv} \tau_{n;j}^{\ast}&=\frac{\tau}{\sqrt{1+\sum_{i,k\in I_j}
\alpha_{n;i}\alpha_{n;k}\sqrt{(1-\omega_{n;i,j}^2)(1-\omega_{n;j,k}^2)}\frac{(\omega_{n;i,k}-\omega_{n;i,j}\omega_{n;j,k})}{\sqrt{(1-\omega_{n;i,j}^2)(1-\omega_{n;k,j}^2)}}}}\\
&\nonumber\to\frac{\tau}{C(\bar{\bLambda},\balpha^{\circ})}\equiv \tau^{\ast}_j
\end{align}
and
\begin{align}
\label{eq:mixed_conv} \alpha_{n;j}^{\ast}\sqrt{\ln n}&=\frac{-\sum_{i\in I_j}\alpha_{n;i}(1-\omega_{n;i,j})\sqrt{\ln n}}{C(\bar{\bOmega}_n,\balpha_n)}\\
&\nonumber\to
\frac{-\sum_{i\in I_j} \lambda_{i,j}^2\alpha_i^{\circ}}{C(\bar{\bLambda},\balpha^{\circ})}.
\end{align}

From \eqref{eq:marg_alpha_conv} and \eqref{eq:marg_ext_conv} we have that as $n\to\infty$
$$
U_i(n/z_i)\approx\Phi_{0,\tau_{i}^{*}}^{\leftarrow}(1-z_i/n),\quad 
U_j(n/z_j)\approx\Phi_{0,\tau_{j}^{*}}^{\leftarrow}(1-z_j/n),
$$
and from \citet[Proposition 0.10]{resnick1987} it follows that as $n\to\infty$
$$
(U_i(n/z_i)-b_{n;i})/a_{n;i} \approx \ln z_i^{-1}, \quad
(U_j(n/z_j)-b_{n;j})/a_{n;j} \approx \ln z_i^{-1}.
$$ 
Then, by
also applying Property Property \ref{lem:prop_esn}\ref{en:second_cond_cond} we obtain
\begin{equation*}\label{eq:stable-tail_1}
\begin{split}
L(\bz)&=\sum_{j=1}^d z_j\lim_{n\to\infty}
\Phi\left\{X_i\leq U_i\left(\frac{n}{z_i}\right),i\in I_j|X_j=U_j\left(\frac{n}{z_j}\right);\tilde{\bOmega}_{n;j},\tilde{\balpha}_{n;j},\tilde{\tau}_{n;j}\right\}\\
&=\sum_{j=1}^d z_j\lim_{n\to\infty}
\Phi\left\{a_{n;i}\left(\frac{U_i\left(\frac{n}{z_i}\right)-b_{n;i}}{a_{n;i}}\right)+b_{n;j}-
\omega_{n;i,j}\left(a_{n;i}\left(\frac{U_i\left(\frac{n}{z_i}\right)-b_{n;j}}{a_{n;i}}\right)+b_{n;j}\right),i\in I_j;\tilde{\bOmega}_{n;j},\tilde{\balpha}_{n;j},\tilde{\tau}_{n;j}\right\}\\
&\approx\sum_{j=1}^d z_j\lim_{n\to\infty}
\Phi\left(\frac{a_{n;i}\ln z_i^{-1}+b_{n;j}-
\omega_{n;i,j}\left(a_{n;j}\ln z_j^{-1}+b_{n;j}\right)}{\{1-\omega_{n;i,j}^2\}^{1/2}},i\in I_j;\bar{\tilde{\bOmega}}_{n;j},\tilde{\balpha}_{n;j},\tilde{\tau}_{n;j}\right).
\end{split}
\end{equation*}
From \eqref{eq:mat_conv} and \eqref{eq:alpha_conv} we have that $\bar{\tilde{\bOmega}}_{n;j}$
converges elementwise to $\bar{\bLambda}_j$ and $\tilde{\balpha}_{n;j}$ converges
componentwise to $\tilde{\balpha}_j$ as $n\to\infty$.
Now, if we assume that $\alpha_{n;i}^{\ast}\geq 0$ and $\alpha_{n;j}^{\ast}<0$, then by
\eqref{eq:marg_alpha_conv}, \eqref{eq:marg_ext_conv} and \eqref{eq:mixed_conv} we have
for all $j\in I$ and $i\in I_j$ that
\begin{align*}
&\frac{a_{n;i}\ln z_i^{-1}+b_{n;j}-
\omega_{n;i,j}\left(a_{n;j}\ln z_j^{-1}+b_{n;j}\right)}{\{1-\omega_{n;i,j}^2\}^{1/2}}\\
&=\frac{\ln z_i^{-1}-\ln z_j^{-1}+(1-\omega_{n;i,j})\ln z_j^{-1}}{\ell_{n}\{1-\omega_{n;i,j}^2\}^{1/2}}+
\frac{\{1-\omega_{n;i,j}\}^{1/2}\ell_{n}}{\{1+\omega_{n;i,j}\}^{1/2}}\\
&-\frac{\ln(2\sqrt{\pi})+(1/2)\ln\ln n  +\ln\Phi\left(\tau_{n;i}^{\ast}/\bar{\alpha}_{n;i}\right)-\ln\Phi\left(\alpha_{n;i}^{\ast}\ell_{n}+\tau_{n;i}^{\ast}\right)}
{\ell_{n}\{1-\omega_{n;i,j}^2\}^{1/2}}\\
&+\omega_{n;i,j}\frac{\ln \sqrt{2\pi}+\ln\Phi\left(\tau_{n;j}^{\ast}/\bar{\alpha}_{n;j}\right)-\ln\Phi\left(\alpha_{n;j}^{\ast}\ell_{n}+\tau_{n;j}^{\ast}\right)\ln\Phi\left(\bar{\alpha}_{n;j}^2\ell_{n}+\alpha_{n;j}^{\ast}\tau_{n;j}^{\ast}\right)}
{\ell_{n}\{1-\omega_{n;i,j}^2\}^{1/2}}\\
&\to \frac{\ln z_i^{-1}-\ln z_j^{-1}}{2 \lambda_{ij}} +\lambda_{ij} +
\frac{\ln \Phi\left(\frac{\tau-\sum_{k\in I_i} \sqrt{2}\lambda^2_{k,i}\alpha^{\circ}_k}
{C(\bar{\bLambda},\balpha^{\circ})}\right)-\ln\Phi\left(\frac{\tau-\sum_{k\in I_j} \sqrt{2}\lambda^2_{k,j}\alpha^{\circ}_k}
{C(\bar{\bLambda},\balpha^{\circ})}\right)}{2\lambda_{i,j}}\\
&-\frac{\ln\Phi\left(\frac{\tau}
{C(\bar{\bLambda},\balpha^{\circ})}\right)-\ln \Phi\left(\frac{\tau}
{C(\bar{\bLambda},\balpha^{\circ})}\right)}{2\lambda_{i,j}}\quad\mbox{as } n\to\infty\\
&=\frac{\ln \tilde{z}_j/\tilde{z}_i}{2\lambda_{i,j}}+\lambda_{i,j}.
\end{align*}
Furthermore we also have 
\begin{align*}
\tilde{\tau}_{n;j}&=\tau-\sum_{i\in I_j}(1-\omega_{n;i,j})\alpha_{n;i}(a_{n;j}\ln z_j^{-1}+b_{n;j})\\
&=\tau - \sum_{i\in I_j}(1-\omega_{n;i,j})\alpha_{n;i}\frac{\ln z_j^{-1}}{\ell_{n}}
-\sum_{i\in I_j}(1-\omega_{n;i,j})\alpha_{n;i}\ell_{n}\\
&+\frac{\sum_{i\in I_j}(1-\omega_{n;i,j})}{\ell_{n}}
\left(
\ln \sqrt{2\pi}+\ln\Phi\left(\tau_{n;j}^{\ast}/\bar{\alpha}_{n;i}\right)-\ln\Phi\left(\alpha_{n;j}^{\ast}\ell_{n}+\tau_{n;j}^{\ast}\right)\ln\Phi\left(\bar{\alpha}_{n;i}^2\ell_{n}+\alpha_{n;j}^{\ast}\tau_{n;j}^{\ast}\right)
\right)\\
&\to \tau + 0 -\sum_{i\in I_j}\sqrt{2}\alpha_i^{\circ}\lambda_{i,j}+0\equiv\tilde{\tau}\quad\mbox{as } n\to\infty.
\end{align*}
Combining all these results together produces the final expression in ($5^*$).
The same results are also obtained with similar steps for the cases ($\alpha_{n;j}^{\ast}\geq 0$ and $\alpha_{n;i}^{\ast}<0$), ($\alpha_{n;i}^{\ast}\geq 0$ and $\alpha_{n;j}^{\ast}\geq0$) and
($\alpha_{n;i}^{\ast}< 0$ and $\alpha_{n;j}^{\ast}<0$).

\bibliographystyle{chicago}
\bibliography{biblio}